\documentclass[aps,amsmath,amssymb,amsfonts,twocolumn]{revtex4}
\usepackage{epsfig}
\usepackage{graphicx}
\usepackage{subfigure}
\usepackage{dcolumn}
\usepackage{bm}
\usepackage{amsthm}
\usepackage{enumitem}
\usepackage{slashed}
\usepackage{braket}
\usepackage{amsmath}
\usepackage{mathtools}

\makeatletter
\def\l@subsubsection#1#2{}
\makeatother

\newcommand{\del}{\partial}

\newcommand{\tr}{\operatorname{tr}}

\newcommand{\diag}{\operatorname{diag}}

\newcommand{\bbR}{\mathbb{R}}

\newcommand{\calA}{\mathcal{A}}

\newcommand{\calF}{\mathcal{F}}
\newcommand{\calG}{\mathcal{G}}

\newcommand{\calO}{\mathcal{O}}

\begin{document}

\title{New diagrammatic framework for higher-spin gravity}

\author{Yasha Neiman}
\email{yashula@icloud.com}
\affiliation{Okinawa Institute of Science and Technology, 1919-1 Tancha, Onna-son, Okinawa 904-0495, Japan}

\date{\today}

\begin{abstract}
We consider minimal type-A higher-spin (HS) gravity in four dimensions, at tree level. We propose new diagrammatic rules for this theory, involving both Fronsdal fields and Didenko-Vasiliev (DV) particles -- linearized versions of HS gravity's ``BPS black hole''. The vertices include a standard minimal coupling between particle and gauge field, the Sleight-Taronna cubic vertex for HS fields, and a recently introduced vertex coupling two HS fields to a DV particle. We show how these ingredients can be combined to reproduce all $n$-point functions of the theory's holographic dual -- the free $O(N)$ vector model. Our diagrammatic rules interpolate between the usual ones of field theory and those of string theory. Our construction can be viewed as a bulk realization of HS algebra.
\end{abstract}

\maketitle

\section{Introduction} \label{sec:intro}

Minimal type-A higher-spin gravity in $D=4$ spacetime dimensions \cite{Vasiliev:1990en,Vasiliev:1995dn,Vasiliev:1999ba} is the interacting theory of an infinite tower of parity-even massless fields, one for each even spin. It is also the conjectured bulk dual \cite{Klebanov:2002ja,Sezgin:2002rt,Sezgin:2003pt,Giombi:2012ms} within AdS/CFT \cite{Maldacena:1997re,Gubser:1998bc,Witten:1998qj,Aharony:1999ti} of a particularly simple boundary theory: the free $O(N)$ vector model of $N$ real scalar fields $\varphi^I$ ($I=1,\dots,N$). Remarkably, this holographic duality can be extended from AdS to de Sitter space \cite{Anninos:2011ui}, thus offering a window into 4d quantum gravity with positive cosmological constant. In the present paper, we consider for simplicity the theory in Euclidean AdS, which we represent as a hyperboloid in flat 5d embedding space $\bbR^{1,4}$:
\begin{align}
 EAdS_4 = \left\{x^\mu\in\bbR^{1,4}\, |\, x_\mu x^\mu = -1, \ x^0 > 0 \right\} \ , \label{eq:EAdS}
\end{align}
where the metric of $\bbR^{1,4}$ is $\eta_{\mu\nu} = \diag(-1,1,1,1,1)$. Since the theory involves infinitely many massless fields interacting at all orders in derivatives, it was always believed to be non-local at distances smaller than the AdS radius. Though exotic, this still implies an expectation of locality at \emph{larger} distances. Recent developments, described below, have challenged this expectation. The goal of this paper is to propose a novel formulation of the theory, in terms of bulk diagrams whose elements' non-locality is confined, as per the original hopes, by $\sim 1$ AdS radius.

A simple formulation of \emph{linearized} HS theory is in terms of Fronsdal fields \cite{Fronsdal:1978rb,Fronsdal:1978vb}: a totally-symmetric rank-$s$ tensor potential for each spin $s$. We encode these as polynomials in an auxiliary vector $u^\mu$:
\begin{align}
 h^{(s)}(x,u) = \frac{1}{s!}\,u^{\mu_1}\!\dots u^{\mu_s}h^{(s)}_{\mu_1\dots\mu_s}(x) \ . \label{eq:h}
\end{align}
In this paper, the potentials \eqref{eq:h} are always traceless, which can be viewed as either a gauge choice or a self-contained framework \cite{Skvortsov:2007kz,Campoleoni:2012th}. An efficient modern approach \cite{Biswas:2002nk,Sleight:2016dba} is to treat both $x$ and $u$ as freely-varying vectors in $\bbR^{1,4}$, with a constraint keeping $h^{(s)}_{\mu_1\dots\mu_s}$ tangential to $EAdS_4$, and a scaling rule that defines its behavior away from $x\cdot x = -1$. Altogether, the constraints on $h^{(s)}(x,u)$ read:
\begin{align}
 (u\cdot \del_u)h^{(s)} &= s h^{(s)} \ ; & (x\cdot\del_u)h^{(s)} &= 0 \ ; \label{eq:h_tensor_properties_1} \\ 
 (x\cdot\del_x)h^{(s)} &= -(s+1) h^{(s)} \ ; & (\del_u\cdot\del_u) h^{(s)} &= 0 \ ,  \label{eq:h_tensor_properties_2}
\end{align}
where $\del^\mu_x$ and $\del_u^\mu$ denote flat $\bbR^{1,4}$ derivatives with respect to the specified vectors. An important example of a free HS field is the boundary-bulk propagator \cite{Mikhailov:2002bp,Costa:2014kfa}:
\begin{align}
 \begin{split}
   &\Pi^{(s)}(x,u;\ell,\lambda) = -\frac{2^s s!(2s+\epsilon)}{4\pi^2(2s)!(s+\epsilon)} \\
   &\qquad \times \frac{\big[(\lambda\cdot x)(\ell\cdot u) - (\ell\cdot x)(\lambda\cdot u)\big]^s}{(\ell\cdot x)^{2s+1}} \ .
 \end{split} \label{eq:Pi}
\end{align}
Here, $\ell^\mu$ is a lightlike vector in $\bbR^{1,4}$ whose direction represents a point on the AdS boundary, $\lambda^\mu$ is a null polarization vector orthogonal to $\ell^\mu$, and $\epsilon=D-4$ is a dimensional regulator that serves to unify the $s=0$ and $s>0$ cases. The propagator \eqref{eq:Pi} obeys the properties \eqref{eq:h_tensor_properties_1}-\eqref{eq:h_tensor_properties_2}, while also being transverse $(\del_u\cdot\del_x)\Pi^{(s)} = 0$. 

In this language, a \emph{cubic vertex} can be written as a differential operator $V^{(s_1,s_2,s_3)}(\del_{x_1},\del_{u_1};\del_{x_2},\del_{u_2};\del_{x_3},\del_{u_3})$ of order $s_i$ in $\del_{u_i}$ ($i=1,2,3$) acting on three fields $h_i^{(s_i)}(x_i,u_i)$, where in the end we set the $x_i$'s equal and integrate over $EAdS_4$:
\begin{align}
    \int_{EAdS_4} d^4x\,V^{(s_1,s_2,s_3)} \prod_{i=1}^3 h^{(s_i)}(x_i,u_i) \Big|_{x_i=x} \ . \label{eq:cubic_vertex}
\end{align} 
When we plug in three propagators \eqref{eq:Pi} into \eqref{eq:cubic_vertex}, we should obtain the boundary 3-point correlator $\left<j^{(s_1)}j^{(s_2)}j^{(s_3)}\right>$ of HS currents. The vertex that accomplishes this is given by a simple formula, found by Sleight and Taronna \cite{Sleight:2016dba}:
\begin{align}
  \begin{split}
     V^{(s_1,s_2,s_3)} ={}& \frac{8\!\left(i\sqrt{2}\right)^{s_1+s_2+s_3}}{\sqrt{N}\,\Gamma(s_1+s_2+s_3+\epsilon)} \\
       &\times \big(Y_{12}^{s_1}Y_{23}^{s_2}Y_{31}^{s_3} + Y_{13}^{s_1}Y_{21}^{s_2}Y_{32}^{s_3}\big) \ , 
  \end{split} \label{eq:V_ST}
\end{align}
where $Y_{ij}\equiv\del_{u_i}\cdot\del_{x_j}$. The symmetrization over the two cyclic structures in \eqref{eq:V_ST} ensures gauge invariance within general traceless gauge \cite{CubicBilocal}. A chiral HS theory based on the self-dual part of the vertex \eqref{eq:V_ST} has been developed in \cite{Ponomarev:2016lrm,Skvortsov:2018jea,Skvortsov:2020wtf,Sharapov:2022awp}.

In contrast to the nice cubic vertex \eqref{eq:V_ST}, the derivation of a \emph{quartic} vertex \cite{Bekaert:2015tva} to reproduce the correlator $\left<j^{(0)}j^{(0)}j^{(0)}j^{(0)}\right>$ yields a result that is \emph{non-local at all scales} \cite{Sleight:2017pcz} (see also \cite{Fotopoulos:2010ay,Taronna:2011kt}). Despite heroic recent efforts in the Vasiliev formalism \cite{Gelfond:2018vmi,Didenko:2018fgx,Didenko:2019xzz,Gelfond:2019tac,Vasiliev:2022med}, this locality problem at the quartic level still stands. In this paper, we propose a resolution to the problem, not just for 4-point functions, but for all $n$-point functions. Specifically, we demonstrate that every $n$-point boundary correlator is equal to a sum of bulk tree diagrams, with \emph{only cubic} vertices that are all local beyond $\sim 1$ AdS radius. Beyond cubic order, our rules for constructing these diagrams are different from those of standard bulk field theory, and bear some resemblance to those of string theory. Note that the sufficiency of \emph{tree} diagrams is a general property of a free boundary dual: the correlators in this case lack $1/N$ corrections, which correspond to loop corrections in the bulk. Of course, for different boundary conditions, the bulk loop corrections will not vanish (and even when they do, one would want to compute this explicitly, as was done in \cite{Giombi:2013fka} for the simplest diagrams).

Our construction proceeds in two steps. The first (section \ref{sec:OPE}) is purely within the boundary theory. There, we introduce a simple yet novel diagrammatic description for the $n$-point correlators, in terms of ``single-trace OPE diagrams''. These are tree diagrams with cubic vertices, and single-trace operators on all the lines. Since single-trace operators correspond to fundamental bulk fields, such diagrams can be interpreted almost directly in the bulk theory. The complication is that the required operators are not just the local HS currents $j^{(s)}_{\mu_1\dots\mu_s}(\ell)$ (dual to the bulk HS fields), but rather the \emph{bilocals}:
\begin{align}
 \calO(\ell,\ell') = \frac{\varphi^I(\ell)\varphi_I(\ell')}{NG(\ell,\ell')} \ , \label{eq:bilocal}
\end{align}
whose Taylor expansion around $\ell=\ell'$ yields the $j^{(s)}_{\mu_1\dots\mu_s}$ and their descendants. Here, the normalization factor $G(\ell,\ell')$ is the propagator of the fundamental boundary fields $\varphi^I$:
\begin{align}
 G(\ell,\ell') = \frac{1}{4\pi\sqrt{-2\ell\cdot\ell'}} \ . \label{eq:G}
\end{align}
Thus, to complete the bulk diagrammatic picture, we need the bulk dual of the bilocals \eqref{eq:bilocal}, as well as bulk expressions for their cubic correlators. These were all recently characterized, and satisfy appropriate bulk locality properties \cite{Neiman:2017mel,David:2020fea,Lysov:2022zlw,CubicBilocal}. In section \ref{sec:rules}, we recall these results, and combine them with the diagrams of section \ref{sec:OPE} to yield the desired bulk rules. A more detailed overview of the recent results \cite{Neiman:2017mel,David:2020fea,Lysov:2022zlw,CubicBilocal} is given in Appendix \ref{app:review}. In Appendix \ref{app:examples}, we illustrate the new bulk rules on examples.

\section{Single-trace OPE diagrams} \label{sec:OPE}

In the boundary theory, all $n$-point correlators of local currents $j^{(s)}$ can be obtained by Taylor-expanding the correlator of $n$ bilocals $\calO(\ell_i,\ell'_i)\equiv \calO_i$. The latter is given by a sum of 1-loop Feynman diagrams:
\begin{align}
  \langle\calO_1\dots\calO_n\rangle = \frac{N^{1-n}}{2n\prod_{p=1}^n\!G(\ell_p,\ell_p')}\sum_{\ell_i\leftrightarrow\ell'_i}\sum_{S_n} \prod_{p=1}^n G(\ell_p',\ell_{p+1}) \label{eq:correlator}
\end{align} 
where the last product is cyclic $\ell_{n+1}\equiv\ell_1$, the inner sum is over the $n!$ permutations of $(\calO_1,\dots,\calO_n)$, and the outer sum is over the $2^n$ permutations between the two endpoints of each $\calO_i$.

Now, a defining property of the boundary propagators $G(\ell,\ell')$ is that their conformal Laplacian w.r.t. each endpoint $\ell,\ell'$ is a boundary delta function:
\begin{align}
 \Box_\ell G(\ell,\ell') = \Box_{\ell'} G(\ell,\ell') = -\delta^3(\ell,\ell') \ . \label{eq:Box_G}
\end{align}
This allows us to ``stitch together'' larger 1-loop diagrams out of smaller ones. In particular, the general correlator \eqref{eq:correlator} can be assembled out of cubic ones, via the following diagrammatic rules:
\begin{enumerate}
	\item Draw an arbitrary trivalent tree graph $\calG$ with $n$ external legs.
	\item Assign one of the $\calO_i$ to each external leg, and a ``dummy'' bilocal $\calO(\ell,\ell')$ to each internal leg.
	\item At each node, compute the cubic correlator of the operators on the 3 surrounding legs.
	\item Integrate each ``dummy'' bilocal's endpoints over the boundary, as:
	\begin{align}
	  \frac{N}{4}\int d^3\ell\,d^3\ell' \big(G(\ell,\ell') \big(\dots\big)\big)\Box_\ell\Box_{\ell'}\big(G(\ell,\ell')\big(\dots\big)\big) \ , \label{eq:bilocal_integral}
	\end{align}
	where ``$\dots$'' are placeholders for the correlator on each side of the internal leg.
	\item Sum over inequivalent permutations of the external legs, multiply by $\calG$'s symmetry factor, and divide by $2n$.
\end{enumerate}
The freedom to choose the graph $\calG$ becomes non-trivial at $n=6$, where two inequivalent  first appear. We show the possible graphs for $n\leq 6$ in figure \ref{fig:graphs} of Appendix \ref{app:examples}. The external bilocals $\calO_i$ can be replaced with local currents $j_i^{(s_i)}$, to produce the standard $n$-point functions. However, the internal-leg integrals \eqref{eq:bilocal_integral} will always feature bilocals. One can think of these rules as constructing OPE diagrams, but with the OPE restricted at every step to single-trace operators, where the integral \eqref{eq:bilocal_integral} acts as a projector onto the space of single-trace bilocals \eqref{eq:bilocal}. The price for this projection is having to sum over permutations of the external legs, which isn't necessary in a standard OPE diagram.

More precisely, the integral \eqref{eq:bilocal_integral} is a projector onto the single-trace sector \emph{times a factor of $\frac{1}{2}$}. That is, upon inserting quadratic single-trace correlators $\langle\calO_1\calO(\ell,\ell')\rangle$ and $\langle\calO(\ell,\ell')\calO_2\rangle$ into the placeholders in \eqref{eq:bilocal_integral}, the integral evaluates to $\frac{1}{2}\langle\calO_1\calO_2\rangle$. One way to see that such a factor of $\frac{1}{2}$ is necessary is to compare with \cite{Sleight:2017pcz}, where a straightforward sum of single-trace projections in different channels produces the quartic correlator with a factor of 2.

\section{The new bulk rules} \label{sec:rules}

We now translate the above ``single-trace OPE'' diagrams into the bulk, to produce the bulk diagrams for $n$-point correlators $\left< j^{(s_1)}\!\dots j^{(s_n)}\right>$. A key step is to identify the bulk dual of the boundary bilocals $\calO(\ell,\ell')$ on the diagram's internal legs. As we review in Appendix \ref{app:review}, this is given by a \emph{geodesic particle worldline} $\gamma(\ell,\ell')$ stretching between the boundary points $\ell,\ell'$ \cite{David:2020fea,Lysov:2022zlw}. This particle produces HS fields $\phi^{(s)}(x,u;\ell,\ell')$ with all spins $s$, which solve Fronsdal's linear field equations with sources on $\gamma(\ell,\ell')$, and form the linearized version of the Didenko-Vasiliev (DV) ``BPS black hole'' \cite{Didenko:2009td,Didenko:2008va}. In our bulk diagrams (Appendix \ref{app:examples}), we depict a DV worldline $\gamma(\ell,\ell')$ as a solid line, and the field $\phi^{(s)}$ as a wavy line emanating from it. The boundary-bulk propagators $\Pi^{(s)}$ corresponding to external currents $j^{(s)}$ are depicted as external wavy lines.

The quadratic correlator of $\calO(\ell,\ell')$ with another (local or bilocal) single-trace boundary operator can be computed as a worldline integral, describing a minimal coupling between the DV worldline of $\calO(\ell,\ell')$ and the bulk field $h^{(s)}(x,u)$ of the second operator (a boundary-bulk propagator $\Pi^{(s)}$ with fixed spin, or a multiplet $\phi^{(s)}$ with all spins):
\begin{align}
 \frac{4}{\sqrt{N}}\sum_s (i\sqrt{2})^s \int_{\gamma(\ell,\ell')} d\tau\,\big(\dot x(\tau)\cdot\del_u\big)^s\,h^{(s)}\big(x(\tau),u\big) \ . \label{eq:2_point}
\end{align}
Here, $\tau$ is the proper time (i.e. length parameter) along $\gamma(\ell,\ell')$, while $x^\mu(\tau),\dot x^\mu(\tau)$ are the corresponding position and 4-velocity (i.e. unit tangent). In bulk diagrams, we depict the coupling \eqref{eq:2_point} as the wavy line depicting $h^{(s)}$ attached to the solid line depicting $\gamma(\ell,\ell')$.

To complete the translation of section \ref{sec:OPE}'s diagrams into the bulk, we need bulk expressions for the cubic correlators $\langle jjj\rangle,\langle jj\calO\rangle,\langle j\calO\calO\rangle,\langle\calO\calO\calO\rangle$. For the $\langle jjj\rangle$ correlator, we use the known cubic vertex \eqref{eq:cubic_vertex}-\eqref{eq:V_ST}, which we depict as usual as a meeting point of three wavy lines. For the cubic correlators involving bilocals, we need two additional kinds of bulk diagram elements \cite{CubicBilocal}. These couple the DV worldline of a bilocal (summing over the possible choices) to the other two operators' bulk fields $h_1^{(s_1)},h_2^{(s_2)}$. The fields are coupled to the worldline either independently -- by multiplying a pair of integrals \eqref{eq:2_point}, each with its worldline-field coupling, or together -- via a single worldline integral, and a new worldline-field-field vertex $V^{(s_1,s_2)}_{\text{new}}\big(\del_{x_1},\del_{u_1};\del_{x_2},\del_{u_2};\dot x(\tau)\big)$:
\begin{align}
   \sum_{s_1,s_2} \int_{\gamma(\ell,\ell')} d\tau\,&V^{(s_1,s_2)}_{\text{new}} h_1^{(s_1)}(x_1,u_1)h_2^{(s_2)}(x_2,u_2)\Big|_{x_1=x_2=x(\tau)} \label{eq:V_new}
\end{align}
In the bulk diagrams, we depict this new vertex as two wavy lines meeting at a solid line. While the formula for $V^{(s_1,s_2)}_{\text{new}}$ is still unknown, it has been established \cite{CubicBilocal} to be local beyond $\sim 1$ AdS radius, as we review in Appendix \ref{app:review}. Thus, the boundary bilocals and their cubic correlators can all be described in terms of local bulk objects. Plugging these into section \ref{sec:OPE}'s diagrams then yields a bulk-local description to all $n$-point functions.

As our final step, we notice a simplification when $h^{(s)}$ in the worldline-field coupling \eqref{eq:2_point} is the field $\phi^{(s)}$ of another bilocal (this will occur in diagrams with $n\geq 5$). Eq. \eqref{eq:2_point} then computes the correlator \eqref{eq:correlator} of two bilocals. When this is acted on in \eqref{eq:bilocal_integral} by boundary Laplacians, eq. \eqref{eq:Box_G} yields delta-functions that set the two bilocals (and thus, their bulk worldlines) equal to each other, with a residual factor of $\frac{1}{2}$ (c.f. the discussion at the end of section \ref{sec:OPE}). This trivialization of some of the bilocal integrals \eqref{eq:bilocal_integral} can be incorporated into the diagrammatic rules, whose final form reads:
\begin{enumerate}
	\item Draw an arbitrary trivalent tree graph $\calG$ with $n$ external legs. Write a boundary-bulk propagator \eqref{eq:Pi} for each external leg.
    \item Draw a solid line across each internal leg of $\calG$. This visually represents a DV worldline, while also highlighting the fact that it splits $\calG$ into two ``sides''. Integrate over the worldline's endpoints as in \eqref{eq:bilocal_integral}, where the $(\dots)$ placeholders correspond to the diagram elements on either ``side'' of the worldline.
    \item Resolve each cubic vertex of $\calG$ into one of the cubic diagrams described above: a usual cubic vertex \eqref{eq:cubic_vertex}-\eqref{eq:V_ST}, a pair of worldline-field couplings \eqref{eq:2_point}, or a wordline-field-field coupling \eqref{eq:V_new}. In all of these, a DV worldline is associated with the DV fields $\phi^{(s)}$, with the spin $s$ summed over.
	\item Sum over all inequivalent diagrams obtained through steps 2-3, sum over inequivalent permutations of the external legs in each diagram, multiply by the symmetry factor of $\calG$, and divide by $2n$.
    \item \emph{After} evaluating the combinatorics as above, we may identify any two worldlines connected by the minimal coupling \eqref{eq:2_point}, replacing one of the associated integrals \eqref{eq:bilocal_integral} by a factor of $\frac{1}{2}$. In this process, some inequivalent orderings of the external legs may become equivalent, i.e. the diagram's symmetry may increase. Note that the ``sides'' of the combined worldline remain unambiguous.
\end{enumerate}
In Appendix \ref{app:examples}, we draw the resulting diagrams for $n\leq 5$, and write out their expressions.

\subsection{Interpreting the integral over worldlines}

Our integral \eqref{eq:bilocal_integral} over DV worldlines is admittedly unusual: instead of integrating over trajectories with fixed endpoints, we integrate over the endpoints while keeping the trajectory a geodesic. The following remarks might make this more palatable. 

The restriction to geodesics may be considered a consequence of HS symmetry, through the requirement that the worldline's HS currents should all be conserved. Indeed, for spin 2, energy-momentum conservation requires the worldline to be geodesic \emph{at leading order in interactions}; it's then plausible to expect that the full HS multiplet will enforce a geodesic at all orders. 

As for the integration over endpoints, it is at least consistent with the standard variational principle for bulk gauge fields (HS or not), which is to hold fixed their \emph{magnetic boundary data}. Our integration over DV worldlines doesn't violate this aspect of the variational principle, because the asymptotic HS fields induced by such worldlines are purely electric \cite{Neiman:2017mel}.

\subsection{Relation to string theory}

In several respects, our diagrammatic rules are intermediate between the Feynman rules for fields and strings. First, the fact that any single ``seed'' graph $\calG$ yields the entire correlator is analogous to how tree-level amplitudes in string theory are given by a single string diagram. Second, the factor of $2n$ is the symmetry factor of an (unoriented) open string diagram with disk topology. Thus, the multiplication by $\calG$'s symmetry factor and division by $2n$ can be seen as ``trading'' the combinatorics of field diagrams for those of string diagrams. Third, the fact that each DV worldline in our diagrams has two ``sides'' is particularly natural if we imagine it as lying on a string worldsheet. Fourth, our use of only cubic vertices can be seen as intermediate between Yang-Mills/GR (which have quartic or higher vertices) and string theory (which has no vertices at all).

Another analogy with string theory is the absence, in a sense, of off-shell fields. In the boundary diagrams of section \ref{sec:OPE}, this manifests as the projection at every step onto the single-trace sector. In our bulk diagrams, the bulk propagators are always attached to a DV worldline, producing the DV solution $\phi^{(s)}$. Thus, instead of arbitrary off-shell fields, we are restricted to the ``almost on-shell'' space of BPS-like DV fields. In fact, as argued in \cite{Lysov:2022zlw}, the DV particle and its fields can be viewed as a ``completion'' of the on-shell field space, in the same way that the string is a completion of its associated spectrum of fields. The holographic intuition behind this view is that the DV particle is dual to the boundary bilocal, whose Taylor expansion forms the tower of HS currents, just as the string is dual to a boundary Wilson loop (or line) \cite{Rey:1998ik,Maldacena:1998im}, whose Taylor expansion similarly forms the tower of local single-trace operators in super-Yang-Mills theory. Another analogy between the DV particle and the string is that the latter can be discovered as one of the BPS solutions of 10d supergravity \cite{Schwarz:1996bh,Blumenhagen:2013fgp}, just like the former is a (linearized) BPS solution of HS gravity \cite{Didenko:2009td}.

\subsection{Relation to HS algebra}

One can view our rules as a bulk-local realization of HS algebra \cite{Fradkin:1986ka} -- the non-commutative product structure $Y_a\star Y_b = Y_a Y_b + iI_{ab}$ behind the infinite-dimensional symmetry group of HS gravity (here, $Y_a$ is a twistor, and $I_{ab}$ is the metric on twistor space). The original cubic vertices found by Fradkin and Vasiliev \cite{Fradkin:1986qy,Fradkin:1987ks} for certain values of spins $(s_1,s_2,s_3)$ were constructed, much like Yang-Mills theory, from the antisymmetric product $\omega_{[\mu}\star\omega_{\nu]}$ acting on the connection master fields $\omega_\mu(x;Y)$. Vasiliev's fully non-linear equations \cite{Vasiliev:1990en,Vasiliev:1995dn,Vasiliev:1999ba} added into the picture a master field $C(x;Y)$ of Weyl curvatures and their derivatives, along with an extra twistor coordinate $Z_a$. In the original approach to the equations, it was found \cite{Boulanger:2015ova} that the remaining cubic vertices (the ones not covered by \cite{Fradkin:1986qy,Fradkin:1987ks}) are given by a structure similar to the \emph{symmetric} product $C\star C$, and that this structure is not consistent with bulk locality. The eventual local formula \eqref{eq:V_ST} for all on-shell cubic vertices was found in \cite{Sleight:2016dba} without any use of HS algebra. 

On the other hand, it was noticed in \cite{Colombo:2012jx,Didenko:2012tv} that the boundary $n$-point functions can all be expressed as traces $\tr_\star(f\star\ldots\star f)$ of symmetrized $\star$-products of a twistor function $f(Y)$, later identified in \cite{Neiman:2017mel} as the \emph{Penrose transform} \cite{Penrose:1986ca,Ward:1990vs} of the bulk field $C(x;Y)$. Moreover, the relevant products $f\star\ldots\star f$ are spanned precisely by the Penrose transforms of linearized DV solutions $\phi^{(s)}$. Now, any symmetrized trace $\tr_\star(f\star\ldots\star f)$ can be constructed from two fundamental operations: the symmetrized product $f\star g + g\star f$ and the pairing $\tr_\star(f\star g)$. This precisely corresponds to our construction above, where the $n$-point correlators are assembled from quadratic and cubic ones, which in turn are composed of bulk-local elements. 

In this view, HS algebra \emph{does} describe local bulk interactions, but one must (a) apply it to the Penrose transform $f(Y)$, and (b) supplement the bulk HS fields with DV particles.

\section{Outlook}

A number of questions are open. First, one should find an explicit expression for the new vertex \eqref{eq:V_new}. Among other benefits, this will allow a direct check of the arguments for its locality made in \cite{CubicBilocal}. Second, it would be good to find a more natural geometric origin for our diagrammatic rules, such as the worldsheet picture of string theory. Finally, our rules should be applied/extended to loop level, which will allow comparison with e.g. the absence of $1/N$ corrections in the free vector model's correlators.

\begin{acknowledgments}
    I am grateful to Adi Armoni, Frederik Denef, Sudip Ghosh, Slava Lysov and Mirian Tsulaia for discussions. This work was supported by the Quantum Gravity Unit of the Okinawa Institute of Science and Technology Graduate University (OIST). The diagrams were drawn using Aidan Sean's tool at \verb|https://www.aidansean.com/feynman/|.
\end{acknowledgments}

\begin{widetext}

\appendix
\section{Review of the Didenko-Vasiliev solution and quadratic/cubic correlators} \label{app:review}

In this Appendix, we review in greater detail the results \cite{Neiman:2017mel,David:2020fea,Lysov:2022zlw,CubicBilocal} relevant for the cubic correlators $\langle jj\calO\rangle,\langle j\calO\calO\rangle,\langle\calO\calO\calO\rangle$ that involve bilocal operators.

\subsection{The linearized solutions}

We begin with the Fronsdal field equations for linearized massless HS fields $h^{(s)}(x,u)$, within the embedding-space formalism described in section \ref{sec:intro}. These are given in terms of the Fronsdal curvature tensor $\calF h^{(s)}(x,u)$ -- an HS generalization of the linearized Ricci tensor. In traceless gauge, the differential operator $\calF$ is given by:
\begin{align}
  \calF = -\del_x\cdot\del_x + \left(u\cdot\del_x + (2s-1)\frac{u\cdot x}{x\cdot x} \right)(\del_u\cdot\del_x) \ . \nonumber
\end{align}
The Fronsdal field equations then read:
\begin{align}
    \left(1 - \frac{1}{4}(g_{\mu\nu}u^\mu u^\nu)(\del_u\cdot\del_u) \right) \calF h^{(s)} = J^{(s)} \ , \label{eq:Fronsdal_eq}
\end{align}
where $g_{\mu\nu} =  \eta_{\mu\nu} - \frac{x_\mu x_\nu}{x\cdot x}$ is the $EAdS_4$ metric at the embedding-space point $x^\mu$. In a straightforward generalization of the Einstein equation, the LHS of \eqref{eq:Fronsdal_eq} is the Fronsdal tensor $\calF h^{(s)}$ with rearranged trace. The source $J^{(s)}(x,u)$ on the RHS of \eqref{eq:Fronsdal_eq} is a totally-symmetric HS current which is double-traceless and conserved up to trace terms:
\begin{align}
     (\del_u\cdot\del_u)^2J^{(s)} = 0 \ ; \quad (\del_u\cdot\nabla_x) J^{(s)} = (g_{\mu\nu}u^\mu u^\nu)(\dots) \ , \nonumber
\end{align}
where the $EAdS_4$ covariant divergence $\del_u\cdot\nabla_x$ is just the flat embedding-space divergence $\del_u\cdot\del_x$ projected back into the tangent space of $EAdS_4$.

The boundary-bulk propagators \eqref{eq:Pi} solve the Fronsdal equation \eqref{eq:Fronsdal_eq} without sources, $J^{(s)}=0$. In contrast, the linearized DV solution $\phi^{(s)}$ solves the equation with sources concentrated on a geodesic worldline $\gamma(\ell,\ell')$:
\begin{align}
     J^{(s)}(x,u) = \frac{Q^{(s)}}{s!}\int_{\gamma(\ell,\ell')} d\tau\,\big(u\cdot\dot x(\tau)\big)^s\,\delta^4\big(x,x(\tau)\big) \ .
\end{align}
Here, $\gamma(\ell,\ell')$ is a geodesic that stretches between a pair of boundary points $\ell,\ell'$:
\begin{align}
   x^\mu(\tau) = \frac{e^\tau \ell^\mu + e^{-\tau}\ell'^\mu}{\sqrt{-2\ell\cdot\ell'}} \ , \label{eq:worldline}
\end{align}
while the charges $Q^{(s)}$ follow a BPS-like proportionality pattern:
\begin{align}
    Q^{(s)} = \frac{4}{\sqrt{N}} \left(i\sqrt{2}\right)^s \ . \label{eq:Q}
\end{align}
The solution $\phi^{(s)}$ itself reads:
\begin{align}
    &\phi^{(s)}(x,u;\ell,\ell') = \frac{1}{\pi\sqrt{N}}\times\frac{1}{R\sqrt{-x\cdot x}} \times \left\{
    \begin{array}{cl}
        1 & \qquad s = 0 \\[0.3em]
        \displaystyle \frac{2}{s!}\left(i\sqrt{2}\right)^s (u\cdot k)^s & \qquad s>0
    \end{array} \right. \ .  \label{eq:phi}
\end{align}
Here, $R$ is the hyperbolic sine of the geodesic distance between $x$ and $\gamma(\ell,\ell')$:
\begin{align}
    R(x;\ell,\ell') &= \sqrt{\frac{2(\ell\cdot x)(\ell'\cdot x)}{(\ell\cdot\ell')(x\cdot x)} - 1} \ ,
\end{align}
while $k^\mu$ is (the Euclidean analog of) an affine tangent to lightrays emanating from $\gamma(\ell,\ell')$ and passing through $x$. It is a null combination of the $EAdS_4$ vectors $t^\mu,r^\mu$ at $x$, which point ``parallel to'' and ``radially away from'' $\gamma(\ell,\ell')$, respectively:
\begin{gather}
   k^\mu(x;\ell,\ell') = \frac{1}{2}\left(t^\mu + \frac{i}{R}r^\mu \right) \ ; \\
   t_\mu(x;\ell,\ell') = \frac{1}{2}\left(\frac{\ell'_\mu}{\ell'\cdot x} - \frac{\ell_\mu}{\ell\cdot x} \right) \ ; \quad
   r_\mu(x;\ell,\ell') = -\frac{x_\mu}{x\cdot x} + \frac{1}{2}\left(\frac{\ell_\mu}{\ell\cdot x} + \frac{\ell'_\mu}{\ell'\cdot x} \right) \ .
\end{gather}
In more diagrammatic terms, the DV solution \eqref{eq:phi} can be thought of as a bulk-bulk propagator $\Pi^{(s)}_{\text{bulk}}(x,u;\tilde x,\tilde u)$, integrated over the worldline \eqref{eq:worldline} through the minimal coupling \eqref{eq:2_point}:
\begin{align}
   \phi^{(s)}(\tilde x,\tilde u;\ell,\ell') = Q^{(s)}\int_{\gamma(\ell,\ell')} d\tau\,\big(\dot x(\tau)\cdot\del_u\big)^s\,\Pi^{(s)}_{\text{bulk}}\big(x(\tau),u;\tilde x,\tilde u\big) \ . \label{eq:minimal_coupling}
\end{align} 
However, since HS bulk-bulk propagators are rather complicated \cite{Costa:2014kfa,Bekaert:2014cea}, especially in traceless gauge, we have opted in this paper to just use the explicit solution \eqref{eq:phi}.

\subsection{Holographic dictionary and quadratic correlators}

The DV solution \eqref{eq:phi} acts as the bulk dual of the boundary bilocal operator \eqref{eq:bilocal}, in two important ways. The first is that, just like the local boundary currents $j^{(s)}$ can be extracted from the Taylor expansion of the bilocal \eqref{eq:bilocal}, so can the boundary-bulk propagator \eqref{eq:Pi} be extracted from the Taylor expansion of \eqref{eq:phi}:
\begin{align}
   D^{(s)}\big[G(\ell,\ell')\calO(\ell,\ell')\big]\Big|_{\ell=\ell'} &= j^{(s)}(\ell,\lambda) \label{eq:D_j} \\
   D^{(s)}\big[G(\ell,\ell')\phi^{(\tilde s)}(x,u;\ell,\lambda)\big]\Big|_{\ell,\ell'} &= \delta_{s,\tilde s}\,\Pi^{(s)}(x,u;\ell,\lambda) \ . \nonumber
\end{align}
Here, $j^{(s)}(\ell,\lambda)$ is a boundary current at the point $\ell$ contracted with the (null, $\ell$-orthogonal) polarization vector $\lambda^\mu$:
\begin{align}
    j^{(s)}(\ell,\lambda) = \lambda^{\mu_1}\!\dots\lambda^{\mu_s} j_{\mu_1\dots\mu_s}(\ell) \ ,
\end{align}
and $D^{(s)}$ is the same differential operator in both cases:
\begin{align}
    D^{(s)}(\del_\ell,\del_{\ell'},\lambda) \equiv \frac{\sqrt{N}\,i^s}{2}\sum_{m=0}^s (-1)^m \binom{2s}{2m} (\lambda\cdot\del_\ell)^m (\lambda\cdot\del_{\ell'})^{s-m} \ . \nonumber
\end{align}
The other (and stronger) sense in which $\phi^{(s)}$ is the bulk dual of $\calO(\ell,\ell')$ concerns its quadratic correlators with other single-trace operators. Specifically, the correlator of $\calO(\ell,\ell')$ with a local $j^{(s)}(L,\lambda)$ or bilocal $\calO(L,L')$ can be evaluated in the bulk by coupling the corresponding fields to the DV worldline $\gamma(\ell,\ell')$ via the same minimal coupling \eqref{eq:minimal_coupling} that defines $\phi^{(s)}$:
\begin{align}
    \left< \calO(\ell,\ell')\,j^{(s)}(L,\lambda) \right> &= Q^{(s)}\int_{\gamma(\ell,\ell')} d\tau\,\big(\dot x(\tau)\cdot\del_u\big)^s\,\Pi^{(s)}\big(x(\tau),u;L,\lambda\big) \ ; \\
    \left< \calO(\ell,\ell')\,\calO(L,L') \right> &= \sum_s Q^{(s)}\int_{\gamma(\ell,\ell')} d\tau\,\big(\dot x(\tau)\cdot\del_u\big)^s\,\phi^{(s)}\big(x(\tau),u;L,L'\big) \ . 
\end{align}

\subsection{Cubic correlators}

We are now ready to consider the simplest cubic correlator involving a bilocal, namely a correlator of the form $\left< j^{(s_1)} j^{(s_2)} \calO\right>$. As discussed in section \ref{sec:rules}, there are two bulk diagrams for this correlator that can be drawn immediately: one with the known cubic vertex \eqref{eq:cubic_vertex}-\eqref{eq:V_ST} coupling the three bulk fields $\Pi^{(s_1)},\Pi^{(s_2)},\phi^{(s)}$, and one with $\Pi^{(s_1)},\Pi^{(s_2)}$ each separately coupled to the bilocal's DV worldline via the minimal coupling \eqref{eq:2_point},\eqref{eq:minimal_coupling}:
\begin{align}
  \begin{split} 
    \calA_{\text{cubic}} ={}& \sum_s\int_{EAdS_4} d^4x\,V^{(s_1,s_2,s)}(\del_{x_1},\del_{u_1};\del_{x_2},\del_{u_2};\del_{x_3},\del_{u_3}) \\
      &\qquad\qquad\qquad\quad \times \Pi^{(s_1)}(x_1,u_1;\ell_1,\lambda_1)\,\Pi^{(s_2)}(x_2,u_2;\ell_2,\lambda_2)\,\phi^{(s)}(x_3,u_3;\ell,\ell') \Big|_{x_1=x_2=x_3=x} \ ; 
  \end{split} \label{eq:A_cubic} \\
  \calA_{\text{minimal}} ={}& \frac{16}{N}\,\prod_{i=1}^2 \left( (i\sqrt{2})^{s_i} \int_{\gamma(\ell,\ell')} d\tau\,\big(\dot x(\tau)\cdot\del_u\big)^{s_i}\,\Pi^{(s_i)}\big(x(\tau),u;\ell_i,\lambda_i\big) \right) \ .
  \label{eq:A_minimal}
\end{align}
Together, these diagrams do not yet capture the full correlator $\left< j^{(s_1)} j^{(s_2)} \calO\right>$. However, the discrepancy can be parameterized as an additional (and still unknown) bulk vertex $V^{(s_1,s_2)}_{\text{new}}(\del_{x_1},\del_{u_1};\del_{x_2},\del_{u_2};\dot x)$ that couples the two propagators $\Pi^{(s_1)},\Pi^{(s_2)}$ to the DV worldline:
\begin{align}
  \begin{split}
    &\left< j^{(s_1)}(\ell_1,\lambda_1)\, j^{(s_2)}(\ell_2,\lambda_2)\,\calO(\ell,\ell') \right> - \big(\calA_{\text{cubic}} + \calA_{\text{minimal}}\big) \\
    &\quad \equiv \int_{\gamma(\ell,\ell')} d\tau\,V^{(s_1,s_2)}_{\text{new}}\big(\del_{x_1},\del_{u_1};\del_{x_2},\del_{u_2};\dot x(\tau)\big)\,
      \Pi^{(s_1)}(x_1,u_1;\ell_1,\lambda_1)\,\Pi^{(s_2)}(x_2,u_2;\ell_2,\lambda_2)\Big|_{x_1=x_2=x(\tau)} \equiv \calA_{\text{new}} \ .
  \end{split} \label{eq:V_new_matching}
\end{align}
As usual in holographic reconstruction, one can always write an ansatz for $V^{(s_1,s_2)}_{\text{new}}$ as a tower of derivatives (with the spacetime symmetries, gauge symmetries and linearized field equations taken into account), and see that the degrees of freedom in this ansatz are just enough to match the correlator as in \eqref{eq:V_new_matching}. The key question is, \emph{does this procedure produce a vertex that is sufficiently local?} As argued in \cite{CubicBilocal}, the answer is yes. This result consists of several building blocks, which we will now review. We begin by splitting the locality requirement in two:
\begin{enumerate}
    \item Radial locality: $V_{\text{new}}$ should not be sensitive to its input fields $h^{(s_1)},h^{(s_2)}$ outside a neighborhood of $\sim 1$ AdS radius from the geodesic.
    \item Time locality: $V_{\text{new}}$ should not be coupling $h^{(s_1)}$ to $h^{(s_2)}$ at points more distant that $\sim 1$ AdS radius \emph{along} the geodesic.
\end{enumerate}
Through arguments detailed in \cite{CubicBilocal}, we can convert these bulk-language requirements into properties of the corresponding diagram \eqref{eq:V_new_matching}, as a function of the boundary points $\ell_1,\ell_2,\ell,\ell'$:
\begin{enumerate}
    \item Radial locality: $\calA_{\text{new}}$ shouldn't have singularities in the ``OPE limit'' $\ell_1=\ell_2$.
    \item Time locality: $\calA_{\text{new}}$ should decay exponentially when $\ell_1,\ell_2$ are separated by a large distance $\Delta\tau$ along $\gamma(\ell,\ell')$, 
\end{enumerate}
Here, in the time-locality criterion, we extended the proper time coordinate $\tau$ on the worldline $\gamma(\ell,\ell')$ to the entire bulk and boundary, by drawing constant-$\tau$ hypersurfaces orthogonal to $\gamma(\ell,\ell')$. Note that, in the bulk, the locality criteria always refer to the IR (large-distance) behavior of the vertex. In the time-locality criterion, this translates directly into IR behavior on the boundary, via the common $\tau$ coordinate. In contrast, in the radial criterion, we see the holographic UV/IR duality at play, converting bulk IR behavior into boundary UV behavior.

Now, it turns out that each of the two locality criteria is satisfied thanks to one of the known diagrams \eqref{eq:A_cubic}-\eqref{eq:A_minimal}. Specifically, $\calA_{\text{cubic}}$ precisely covers the part of the correlator that would require a radially non-local $V_{\text{new}}$, while $\calA_{\text{minimal}}$ precisely covers the part that would require a time-non-local $V_{\text{new}}$. 

The radial-locality result of \cite{CubicBilocal} consists of several steps:
\begin{itemize}
    \item Since $\calA_{\text{minimal}}$ by itself is radially-local, we focus on the difference $\left<j^{(s_1)} j^{(s_2)} j^{(s)}\right> - \calA_{\text{cubic}}$.
    \item We notice that, if one cares only about the correlators of $\calO(\ell,\ell')$ with boundary operators \emph{outside of} a region that includes $\ell,\ell'$, then $\calO(\ell,\ell')$ can be replaced with a superposition of local currents $j^{(s)}$ inside the region.
    \item We recall that the standard cubic vertex $V^{(s_1,s_2,s)}$, acting on boundary-bulk propagators $\Pi^{(s_1)},\Pi^{(s_2)},\Pi^{(s)}$, precisely computes the resulting correlators $\left<j^{(s_1)} j^{(s_2)} j^{(s)}\right>$.
    \item We notice that, in any region that \emph{doesn't include} the DV worldline, the bulk DV field $\phi^{(s)}$ is free, and can thus be expressed as a superposition of boundary-bulk propagators $\Pi^{(s)}$, up to gauge.
    \item It can be proved that the symmetrized form \eqref{eq:V_ST} of the cubic vertex $V^{(s_1,s_2,s)}$ is sufficiently gauge-invariant such that $\calA_{\text{cubic}}$ is insensitive to the above difference in gauge.
    \item The upshot is that the difference $\left<j^{(s_1)} j^{(s_2)} j^{(s)}\right> - \calA_{\text{cubic}}$ can be localized to a bulk region distant from $\ell_1$ and $\ell_2$. This leads to the desired analyticity in the $\ell_1=\ell_2$ limit. 
\end{itemize}
The time-locality result is technically simpler, yet is more tightly linked with our introduction of DV worldlines as a fundamental bulk object. The result can be summarized simply as:
\begin{itemize}
    \item Since $\calA_{\text{cubic}}$ by itself is radially-local, we focus on the difference $\left<j^{(s_1)} j^{(s_2)} j^{(s)}\right> - \calA_{\text{minimal}}$.
    \item Being the difference between a cubic bulk correlator and a product of quadratic correlators, this can be evaluated explicitly. We find a cancellation of terms that \emph{grow} exponentially with $\Delta\tau$, while the remaining terms \emph{decrease} exponentially, as required.
\end{itemize}
The cancellation mentioned above requires a particular normalization of $\calO(\ell,\ell')$, which fixes the overall coefficient in \eqref{eq:bilocal} and \eqref{eq:Q}-\eqref{eq:phi}. 

This concludes our treatment of the correlator $\langle jj\calO\rangle$. By arguments similar to those we used for radial locality (i.e. dividing the bulk into regions and invoking gauge-invariance), its can be shown that the same diagrammatic elements suffice for the remaining correlators $\langle j\calO\calO\rangle$,$\langle\calO\calO\calO\rangle$. The general cubic correlator can be written as:
\begin{align}
  \begin{split}
    &\sum_{s_1,s_2,s_3} \int_{EAdS_4} d^4x\,V^{(s_1,s_2,s_3)}h_1^{(s_1)}h_2^{(s_2)}h_3^{(s_3)} 
       + \sum_i\prod_{j\neq i}\left( \sum_{s_j} Q^{(s_j)} \int_{\gamma(\ell_i,\ell'_i)} d\tau\,\big(\dot x(\tau)\cdot\del_u\big)^{s_j}\,h_j^{(s_j)}\big(x(\tau),u\big) \right) \\
       &\quad + \sum_i\sum_{s_j,s_k} \int_{\gamma(\ell_i,\ell'_i)} d\tau\,V^{(s_j,s_k)}_{\text{new}}h_j^{(s_j)}h_k^{(s_k)} \ .
  \end{split}
\end{align}
Here, $h_1^{(s_1)},h_2^{(s_2)},h_3^{(s_3)}$ are the bulk fields corresponding to the three boundary operators, and the label $i$ runs over any bilocal operators among the three. In the first line, the label $j$ runs over the two operators other than chosen $i$'th bilocal; in the second line, these two other operators are denoted by the labels $j,k$. 

This concludes our summary of the bulk structure and locality properties of cubic correlators, which were invoked in the beginning of section \ref{sec:rules} in order to translate section \ref{sec:OPE}'s new boundary diagrams into the bulk.

\section{Examples illustrating the new diagrammatic rules} \label{app:examples}

In this Appendix, we provide examples for the diagrams generated by the rules in the main text. Firstly, we illustrate in figure \ref{fig:graphs} the trivalent graphs $\calG$ from section \ref{sec:OPE}, for $3\leq n\leq 6$ external legs. Starting from $n=6$, there is more than one choice of $\calG$. However, any single choice computes the entire $n$-point correlator. 
\begin{figure*}%
    \centering%
    \includegraphics[scale=0.35]{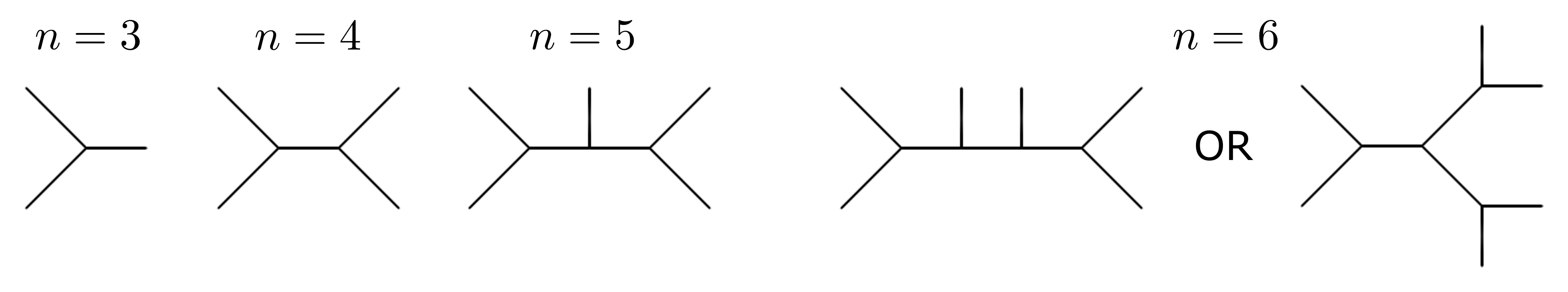} \\
    \caption{Trivalent graphs $\calG$ for $n\leq 6$ external legs. Starting from $n=6$, there's more than one choice of $\calG$, but each choice computes the full correlator.}
    \label{fig:graphs} 
\end{figure*}

Next, we restrict attention to $3\leq n\leq 5$ (where there is only one option for the graph $\calG$), and detail the bulk diagrams that arise from the rules of section \ref{sec:rules}. 

\subsection{$n=3$}

\noindent Here, the symmetry factor of $\calG$ is $S = 3! = 6$. Thus, the combinatorial factor multiplying the diagrams is:
\begin{align}
    \frac{S}{2n} = \frac{6}{2\cdot 3} = 1 \ .
\end{align} 
The only diagram is: $\vcenter{\includegraphics[scale=0.2]{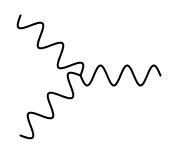}}$
It has just one inequivalent ordering of the external legs. Thus, we recover the standard bulk expression for the cubic correlator:
\begin{align}
    \left< j^{(s_1)}(\ell_1,\lambda_1)\,j^{(s_2)}(\ell_2,\lambda_2)\,j^{(s_3)}(\ell_3,\lambda_3)\right> 
      ={}& \int_{EAdS_4} d^4x\,V^{(s_1,s_2,s_3)}(\del_{x_1},\del_{u_1};\del_{x_2},\del_{u_2};\del_{x_3},\del_{u_3}) \\ 
    &\times \Pi^{(s_1)}(x_1,u_1;\ell_1,\lambda_1)\,\Pi^{(s_2)}(x_2,u_2;\ell_2,\lambda_2)\,\Pi^{(s_3)}(x_3,u_3;\ell_3,\lambda_3) \Big|_{x_1=x_2=x_3=x} \ . \nonumber
\end{align}

\subsection{$n=4$}

\noindent Here, the symmetry factor of $\calG$ is $S = 2^3 = 8$. Thus, the combinatorial factor multiplying the diagrams is again:
\begin{align}
   \frac{S}{2n} = \frac{8}{2\cdot 4} = 1 \ .  
\end{align}
There are 6 bulk diagrams, each containing one DV worldline. The individual diagrams are as follows.

\noindent Diagram no. 1: $\vcenter{\includegraphics[scale=0.2]{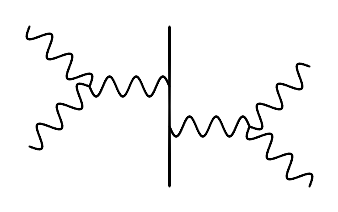}}$
It has $\binom{4}{2}/2 = 3$ inequivalent orderings of the external legs. The expression for one sample ordering reads:
\begin{align}
    \begin{split}
       &\frac{N}{4}\int d^3\ell\,d^3\ell'\,G(\ell,\ell')\sum_{s_5}\int_{EAdS_4} d^4x_{\text{left}}\,V^{(s_1,s_2,s_5)}(\del_{x_1},\del_{u_1};\del_{x_2},\del_{u_2};\del_{x_5},\del_{u_5}) \\
       &\qquad\qquad\qquad\qquad \times \Pi^{(s_1)}(x_1,u_1;\ell_1,\lambda_1)\,\Pi^{(s_2)}(x_2,u_2;\ell_2,\lambda_2)\,\phi^{(s_5)}(x_5,u_5;\ell,\ell') \Big|_{x_1=x_2=x_5=x_{\text{left}}} \\
       &\quad \times \Box_\ell\Box_{\ell'}\,G(\ell,\ell') \sum_{s_6}\int_{EAdS_4} d^4x_{\text{right}}\,V^{(s_3,s_4,s_6)}(\del_{x_3},\del_{u_3};\del_{x_4},\del_{u_4};\del_{x_6},\del_{u_6}) \\
       &\qquad\qquad\qquad\qquad \times \Pi^{(s_3)}(x_3,u_3;\ell_3,\lambda_3)\,\Pi^{(s_4)}(x_4,u_4;\ell_4,\lambda_4)\,\phi^{(s_6)}(x_6,u_6;\ell,\ell') \Big|_{x_3=x_4=x_6=x_{\text{right}}} \ ,
    \end{split}
\end{align}
where the boundary Laplacians $\Box_\ell\Box_{\ell'}$ act on everything to their right.

\noindent Diagram no. 2: $\vcenter{\includegraphics[scale=0.2]{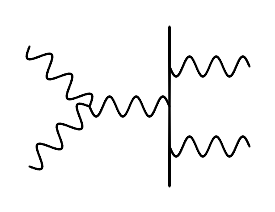}}$
It has $\binom{4}{2}=6$ inequivalent orderings of the external legs. The expression for one sample ordering reads:
\begin{align}
    \begin{split}
        &\frac{N}{4}\int d^3\ell\,d^3\ell'\,G(\ell,\ell')\sum_{s_5}\int_{EAdS_4} d^4x\,V^{(s_1,s_2,s_5)}(\del_{x_1},\del_{u_1};\del_{x_2},\del_{u_2};\del_{x_5},\del_{u_5}) \\
        &\qquad\qquad\qquad\qquad \times \Pi^{(s_1)}(x_1,u_1;\ell_1,\lambda_1)\,\Pi^{(s_2)}(x_2,u_2;\ell_2,\lambda_2)\,\phi^{(s_5)}(x_5,u_5;\ell,\ell') \Big|_{x_1=x_2=x_5=x} \\
        &\quad \times \Box_\ell\Box_{\ell'}\,G(\ell,\ell')\,Q^{(s_3)}\int_{\gamma(\ell,\ell')}  d\tau_3\,\big(\dot x(\tau_3)\cdot\del_{u_3}\big)^{s_3}\, \Pi^{(s_3)}\big(x(\tau_3),u_3;\ell_3,\lambda_3\big) \\
        &\qquad\qquad \times Q^{(s_4)}\int_{\gamma(\ell,\ell')}  d\tau_4\,\big(\dot x(\tau_4)\cdot\del_{u_4}\big)^{s_4}\, \Pi^{(s_4)}\big(x(\tau_4),u_4;\ell_4,\lambda_4\big) \ .
    \end{split}
\end{align}
Diagram no. 3: $\vcenter{\includegraphics[scale=0.2]{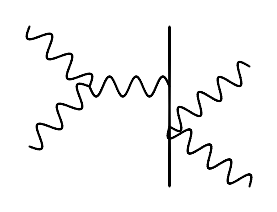}}$
It has $\binom{4}{2}=6$ inequivalent orderings of the external legs. The expression for one sample ordering reads:
\begin{align}
    \begin{split}
        &\frac{N}{4}\int d^3\ell\,d^3\ell'\,G(\ell,\ell')\sum_{s_5}\int_{EAdS_4} d^4x\,V^{(s_1,s_2,s_5)}(\del_{x_1},\del_{u_1};\del_{x_2},\del_{u_2};\del_{x_5},\del_{u_5}) \\
        &\qquad\qquad\qquad\qquad \times \Pi^{(s_1)}(x_1,u_1;\ell_1,\lambda_1)\,\Pi^{(s_2)}(x_2,u_2;\ell_2,\lambda_2)\,\phi^{(s_5)}(x_5,u_5;\ell,\ell') \Big|_{x_1=x_2=x_5=x} \\
        &\quad \times \Box_\ell\Box_{\ell'}\,G(\ell,\ell') \int_{\gamma(\ell,\ell')}  d\tau\,V^{(s_3,s_4)}_{\text{new}}\big(\del_{x_3},\del_{u_3};\del_{x_4},\del_{u_4};\dot x(\tau)\big) \\
        &\qquad\qquad\qquad\qquad \times \Pi^{(s_3)}(x_3,u_3;\ell_3,\lambda_3)\,\Pi^{(s_4)}(x_4,u_4;\ell_4,\lambda_4) \Big|_{x_3=x_4=x(\tau)} \ .
    \end{split}
\end{align}
Diagram no. 4: $\vcenter{\includegraphics[scale=0.2]{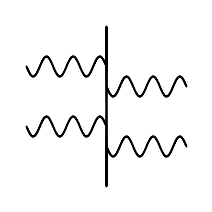}}$
It has $\binom{4}{2}/2 = 3$ inequivalent orderings of the external legs. The expression for one sample ordering reads:
\begin{align}
    \begin{split}
        &\frac{N}{4}\int d^3\ell\,d^3\ell'\,G(\ell,\ell')\,Q^{(s_1)}\int_{\gamma(\ell,\ell')}  d\tau_1\,\big(\dot x(\tau_1)\cdot\del_{u_1}\big)^{s_1}\, \Pi^{(s_1)}\big(x(\tau_1),u_1;\ell_1,\lambda_1\big) \\
        &\qquad\qquad \times Q^{(s_2)}\int_{\gamma(\ell,\ell')}  d\tau_2\,\big(\dot x(\tau_2)\cdot\del_{u_2}\big)^{s_2}\, \Pi^{(s_2)}\big(x(\tau_2),u_2;\ell_2,\lambda_2\big) \\
        &\quad \times \Box_\ell\Box_{\ell'}\,G(\ell,\ell')\,Q^{(s_3)}\int_{\gamma(\ell,\ell')}  d\tau_3\,\big(\dot x(\tau_3)\cdot\del_{u_3}\big)^{s_3}\, \Pi^{(s_3)}\big(x(\tau_3),u_3;\ell_3,\lambda_3\big) \\
        &\qquad\qquad \times Q^{(s_4)}\int_{\gamma(\ell,\ell')}  d\tau_4\,\big(\dot x(\tau_4)\cdot\del_{u_4}\big)^{s_4}\, \Pi^{(s_4)}\big(x(\tau_4),u_4;\ell_4,\lambda_4\big) \ .
    \end{split}
\end{align}
Diagram no. 5: $\vcenter{\includegraphics[scale=0.2]{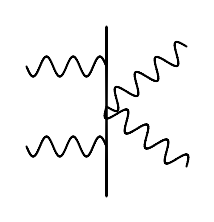}}$
It has $\binom{4}{2} = 6$ inequivalent orderings of the external legs. The expression for one sample ordering reads:
\begin{align}
    \begin{split}
        &\frac{N}{4}\int d^3\ell\,d^3\ell'\,G(\ell,\ell')\,Q^{(s_1)}\int_{\gamma(\ell,\ell')}  d\tau_1\,\big(\dot x(\tau_1)\cdot\del_{u_1}\big)^{s_1}\, \Pi^{(s_1)}\big(x(\tau_1),u_1;\ell_1,\lambda_1\big) \\
        &\qquad\qquad \times Q^{(s_2)}\int_{\gamma(\ell,\ell')}  d\tau_2\,\big(\dot x(\tau_2)\cdot\del_{u_2}\big)^{s_2}\, \Pi^{(s_2)}\big(x(\tau_2),u_2;\ell_2,\lambda_2\big) \\
        &\quad \times \Box_\ell\Box_{\ell'}\,G(\ell,\ell')\int_{\gamma(\ell,\ell')}  d\tau\,V^{(s_3,s_4)}_{\text{new}}\big(\del_{x_3},\del_{u_3};\del_{x_4},\del_{u_4};\dot x(\tau)\big) \\
        &\qquad\qquad\qquad\qquad \times \Pi^{(s_3)}(x_3,u_3;\ell_3,\lambda_3)\,\Pi^{(s_4)}(x_4,u_4;\ell_4,\lambda_4) \Big|_{x_3=x_4=x(\tau)} \ .
    \end{split}
\end{align}
Diagram no. 6: $\vcenter{\includegraphics[scale=0.2]{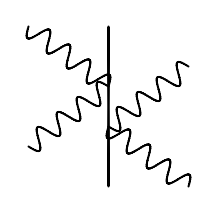}}$
It has $\binom{4}{2}/2 = 3$ inequivalent orderings of the external legs. The expression for one sample ordering reads:
\begin{align}
    \begin{split}
        &\frac{N}{4}\int d^3\ell\,d^3\ell'\,G(\ell,\ell')\int_{\gamma(\ell,\ell')}  d\tau_{\text{left}}\,V^{(s_1,s_2)}_{\text{new}}\big(\del_{x_1},\del_{u_1};\del_{x_2},\del_{u_2};\dot x(\tau_{\text{left}})\big) \\
        &\qquad\qquad\qquad\qquad \times \Pi^{(s_1)}(x_1,u_1;\ell_1,\lambda_1)\,\Pi^{(s_2)}(x_2,u_2;\ell_2,\lambda_2) \Big|_{x_1=x_2=x(\tau_{\text{left}})} \\
        &\quad \times \Box_\ell\Box_{\ell'}\,G(\ell,\ell')\int_{\gamma(\ell,\ell')}  d\tau_{\text{right}}\,V^{(s_3,s_4)}_{\text{new}}\big(\del_{x_3},\del_{u_3};\del_{x_4},\del_{u_4};\dot x(\tau_{\text{right}})\big) \\
        &\qquad\qquad\qquad\qquad \times \Pi^{(s_3)}(x_3,u_3;\ell_3,\lambda_3)\,\Pi^{(s_4)}(x_4,u_4;\ell_4,\lambda_4) \Big|_{x_3=x_4=x(\tau_{\text{right}})} \ .
    \end{split}
\end{align}

\subsection{$n=5$}

\noindent Here, the symmetry factor of $\calG$ is again $S = 2^3 = 8$. Thus, the combinatorial factor multiplying the diagrams is:
\begin{align}
    \frac{S}{2n} = \frac{8}{2\cdot 5} = \frac{4}{5} \ .  
\end{align}
There are 24 bulk diagrams. Of these, the first 15 contain two DV worldlines. In the other 9, the two worldlines can be identified into one. The individual diagrams are as follows.

\noindent Diagram no. 1: $\vcenter{\includegraphics[scale=0.2]{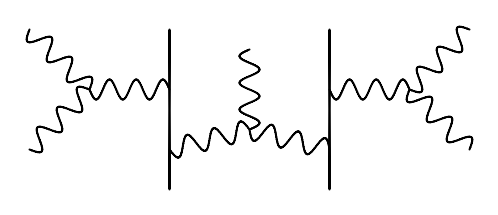}}$
It has $5\cdot\binom{4}{2}/2 = 15$ inequivalent orderings of the external legs. The expression for one sample ordering reads:
\begin{align}
    \begin{split}
        &\left(\frac{N}{4}\right)^2 \int d^3\ell\,d^3\ell'\,G(\ell,\ell')\sum_{s_6}\int_{EAdS_4} d^4x_{\text{left}}\,V^{(s_1,s_2,s_6)}(\del_{x_1},\del_{u_1};\del_{x_2},\del_{u_2};\del_{x_6},\del_{u_6}) \\
        &\qquad\qquad\qquad\qquad \times \Pi^{(s_1)}(x_1,u_1;\ell_1,\lambda_1)\,\Pi^{(s_2)}(x_2,u_2;\ell_2,\lambda_2)\,\phi^{(s_6)}(x_6,u_6;\ell,\ell') \Big|_{x_1=x_2=x_6=x_{\text{left}}} \\
        &\quad \times \Box_\ell\Box_{\ell'}\,G(\ell,\ell')\int d^3\tilde\ell\,d^3\tilde\ell'\,G(\tilde\ell,\tilde\ell')\sum_{s_7,s_8}\int_{EAdS_4}
           d^4x_{\text{mid}}\,V^{(s_3,s_7,s_8)}(\del_{x_3},\del_{u_3};\del_{x_7},\del_{u_7};\del_{x_8},\del_{u_8}) \\
        &\qquad\qquad\qquad\qquad \times \Pi^{(s_3)}(x_3,u_3;\ell_3,\lambda_3)\,\phi^{(s_7)}(x_7,u_7;\ell,\ell')\,\phi^{(s_8)}(x_8,u_8;\tilde\ell,\tilde\ell') \Big|_{x_3=x_7=x_8=x_{\text{mid}}} \\
        &\quad \times \Box_{\tilde\ell}\Box_{\tilde\ell'}\,G(\tilde\ell,\tilde\ell') \sum_{s_9}\int_{EAdS_4} d^4x_{\text{right}}\,V^{(s_4,s_5,s_9)}(\del_{x_4},\del_{u_4};\del_{x_5},\del_{u_5};\del_{x_9},\del_{u_9}) \\
        &\qquad\qquad\qquad\qquad \times \Pi^{(s_4)}(x_4,u_4;\ell_4,\lambda_4)\,\Pi^{(s_5)}(x_5,u_5;\ell_5,\lambda_5)\,\phi^{(s_9)}(x_9,u_9;\tilde\ell,\tilde\ell') \Big|_{x_4=x_5=x_9=x_{\text{right}}} \ .
    \end{split}
\end{align}
Diagram no. 2: $\vcenter{\includegraphics[scale=0.2]{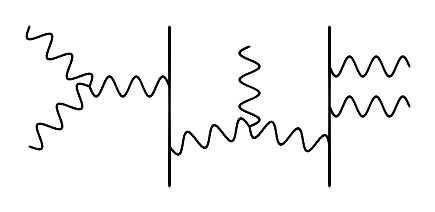}}$
It has $5\cdot\binom{4}{2} = 30$ inequivalent orderings of the external legs. The expression for one sample ordering reads:
\begin{align}
    \begin{split}
        &\left(\frac{N}{4}\right)^2 \int d^3\ell\,d^3\ell'\,G(\ell,\ell')\sum_{s_6}\int_{EAdS_4} d^4x_{\text{left}}\,V^{(s_1,s_2,s_6)}(\del_{x_1},\del_{u_1};\del_{x_2},\del_{u_2};\del_{x_6},\del_{u_6}) \\
        &\qquad\qquad\qquad\qquad \times \Pi^{(s_1)}(x_1,u_1;\ell_1,\lambda_1)\,\Pi^{(s_2)}(x_2,u_2;\ell_2,\lambda_2)\,\phi^{(s_6)}(x_6,u_6;\ell,\ell') \Big|_{x_1=x_2=x_6=x_{\text{left}}} \\
        &\quad \times \Box_\ell\Box_{\ell'}\,G(\ell,\ell')\int d^3\tilde\ell\,d^3\tilde\ell'\,G(\tilde\ell,\tilde\ell')\sum_{s_7,s_8}\int_{EAdS_4}
        d^4x_{\text{mid}}\,V^{(s_3,s_7,s_8)}(\del_{x_3},\del_{u_3};\del_{x_7},\del_{u_7};\del_{x_8},\del_{u_8}) \\
        &\qquad\qquad\qquad\qquad \times \Pi^{(s_3)}(x_3,u_3;\ell_3,\lambda_3)\,\phi^{(s_7)}(x_7,u_7;\ell,\ell')\,\phi^{(s_8)}(x_8,u_8;\tilde\ell,\tilde\ell') \Big|_{x_3=x_7=x_8=x_{\text{mid}}} \\
        &\quad \times \Box_{\tilde\ell}\Box_{\tilde\ell'}\,G(\tilde\ell,\tilde\ell')\,Q^{(s_4)}\int_{\gamma(\tilde\ell,\tilde\ell')}  d\tau_4\,\big(\dot x(\tau_4)\cdot\del_{u_4}\big)^{s_4}\,
        \Pi^{(s_4)}\big(x(\tau_4),u_4;\ell_4,\lambda_4\big) \\
        &\qquad\qquad \times Q^{(s_5)}\int_{\gamma(\tilde\ell,\tilde\ell')}  d\tau_5\,\big(\dot x(\tau_5)\cdot\del_{u_5}\big)^{s_5}\,\Pi^{(s_5)}\big(x(\tau_5),u_5;\ell_5,\lambda_5\big) \ .
    \end{split}
\end{align}
Diagram no. 3: $\vcenter{\includegraphics[scale=0.2]{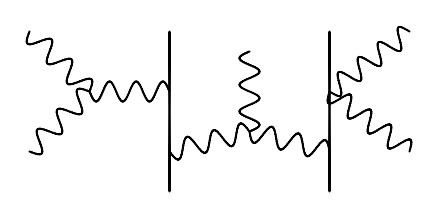}}$
It has $5\cdot\binom{4}{2} = 30$ inequivalent orderings of the external legs. The expression for one sample ordering reads:
\begin{align}
    \begin{split}
        &\left(\frac{N}{4}\right)^2 \int d^3\ell\,d^3\ell'\,G(\ell,\ell')\sum_{s_6}\int_{EAdS_4} d^4x_{\text{left}}\,V^{(s_1,s_2,s_6)}(\del_{x_1},\del_{u_1};\del_{x_2},\del_{u_2};\del_{x_6},\del_{u_6}) \\
        &\qquad\qquad\qquad\qquad \times \Pi^{(s_1)}(x_1,u_1;\ell_1,\lambda_1)\,\Pi^{(s_2)}(x_2,u_2;\ell_2,\lambda_2)\,\phi^{(s_6)}(x_6,u_6;\ell,\ell') \Big|_{x_1=x_2=x_6=x_{\text{left}}} \\
        &\quad \times \Box_\ell\Box_{\ell'}\,G(\ell,\ell')\int d^3\tilde\ell\,d^3\tilde\ell'\,G(\tilde\ell,\tilde\ell')\sum_{s_7,s_8}\int_{EAdS_4}
        d^4x_{\text{mid}}\,V^{(s_3,s_7,s_8)}(\del_{x_3},\del_{u_3};\del_{x_7},\del_{u_7};\del_{x_8},\del_{u_8}) \\
        &\qquad\qquad\qquad\qquad \times \Pi^{(s_3)}(x_3,u_3;\ell_3,\lambda_3)\,\phi^{(s_7)}(x_7,u_7;\ell,\ell')\,\phi^{(s_8)}(x_8,u_8;\tilde\ell,\tilde\ell') \Big|_{x_3=x_7=x_8=x_{\text{mid}}} \\
        &\quad \times \Box_{\tilde\ell}\Box_{\tilde\ell'}\,G(\tilde\ell,\tilde\ell') \int_{\gamma(\tilde\ell,\tilde\ell')}  d\tau\,V^{(s_4,s_5)}_{\text{new}}\big(\del_{x_4},\del_{u_4};\del_{x_5},\del_{u_5};\dot x(\tau)\big) \\
        &\qquad\qquad\qquad\qquad \times \Pi^{(s_4)}(x_4,u_4;\ell_4,\lambda_4)\,\Pi^{(s_5)}(x_5,u_5;\ell_5,\lambda_5) \Big|_{x_4=x_5=x(\tau)} \ .
    \end{split}
\end{align}
Diagram no. 4: $\vcenter{\includegraphics[scale=0.2]{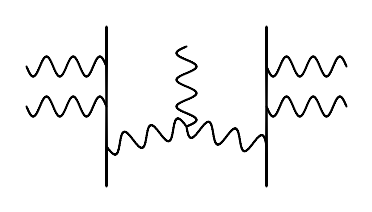}}$
It has $5\cdot\binom{4}{2}/2 = 15$ inequivalent orderings of the external legs. The expression for one sample ordering reads:
\begin{align}
    \begin{split}
        &\left(\frac{N}{4}\right)^2 \int d^3\ell\,d^3\ell'\,G(\ell,\ell')\,Q^{(s_1)}\int_{\gamma(\ell,\ell')} d\tau_1\,\big(\dot x(\tau_1)\cdot\del_{u_1}\big)^{s_1}\,\Pi^{(s_1)}\big(x(\tau_1),u_1;\ell_1,\lambda_1\big) \\
        &\qquad\qquad \times Q^{(s_2)}\int_{\gamma(\ell,\ell')} d\tau_2\,\big(\dot x(\tau_2)\cdot\del_{u_2}\big)^{s_2}\,\Pi^{(s_2)}\big(x(\tau_2),u_2;\ell_2,\lambda_2\big) \\
        &\quad \times \Box_\ell\Box_{\ell'}\,G(\ell,\ell')\int d^3\tilde\ell\,d^3\tilde\ell'\,G(\tilde\ell,\tilde\ell')\sum_{s_6,s_7}\int_{EAdS_4}
        d^4x\,V^{(s_3,s_6,s_7)}(\del_{x_3},\del_{u_3};\del_{x_6},\del_{u_6};\del_{x_7},\del_{u_7}) \\
        &\qquad\qquad\qquad\qquad \times \Pi^{(s_3)}(x_3,u_3;\ell_3,\lambda_3)\,\phi^{(s_6)}(x_6,u_6;\ell,\ell')\,\phi^{(s_7)}(x_7,u_7;\tilde\ell,\tilde\ell') \Big|_{x_3=x_6=x_7=x} \\
        &\quad \times \Box_{\tilde\ell}\Box_{\tilde\ell'}\,G(\tilde\ell,\tilde\ell')\,Q^{(s_4)}\int_{\gamma(\tilde\ell,\tilde\ell')} d\tau_4\,\big(\dot x(\tau_4)\cdot\del_{u_4}\big)^{s_4}\,
        \Pi^{(s_4)}\big(x(\tau_4),u_4;\ell_4,\lambda_4\big) \\
        &\qquad\qquad \times Q^{(s_5)}\int_{\gamma(\tilde\ell,\tilde\ell')} d\tau_5\,\big(\dot x(\tau_5)\cdot\del_{u_5}\big)^{s_5}\,\Pi^{(s_5)}\big(x(\tau_5),u_5;\ell_5,\lambda_5\big) \ .
    \end{split}
\end{align}
Diagram no. 5: $\vcenter{\includegraphics[scale=0.2]{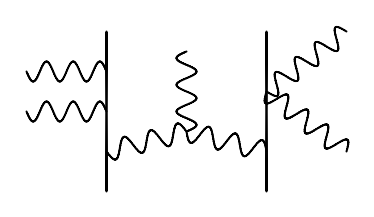}}$
It has $5\cdot\binom{4}{2} = 30$ inequivalent orderings of the external legs. The expression for one sample ordering reads:
\begin{align}
    \begin{split}
        &\left(\frac{N}{4}\right)^2 \int d^3\ell\,d^3\ell'\,G(\ell,\ell')\,Q^{(s_1)}\int_{\gamma(\ell,\ell')}  d\tau_1\,\big(\dot x(\tau_1)\cdot\del_{u_1}\big)^{s_1}\,\Pi^{(s_1)}\big(x(\tau_1),u_1;\ell_1,\lambda_1\big) \\
        &\qquad\qquad \times Q^{(s_2)}\int_{\gamma(\ell,\ell')}  d\tau_2\,\big(\dot x(\tau_2)\cdot\del_{u_2}\big)^{s_2}\,\Pi^{(s_2)}\big(x(\tau_2),u_2;\ell_2,\lambda_2\big) \\
        &\quad \times \Box_\ell\Box_{\ell'}\,G(\ell,\ell')\int d^3\tilde\ell\,d^3\tilde\ell'\,G(\tilde\ell,\tilde\ell')\sum_{s_6,s_7}\int_{EAdS_4}
        d^4x\,V^{(s_3,s_6,s_7)}(\del_{x_3},\del_{u_3};\del_{x_6},\del_{u_6};\del_{x_7},\del_{u_7}) \\
        &\qquad\qquad\qquad\qquad \times \Pi^{(s_3)}(x_3,u_3;\ell_3,\lambda_3)\,\phi^{(s_6)}(x_6,u_6;\ell,\ell')\,\phi^{(s_7)}(x_7,u_7;\tilde\ell,\tilde\ell') \Big|_{x_3=x_6=x_7=x} \\
        &\quad \times \Box_{\tilde\ell}\Box_{\tilde\ell'}\,G(\tilde\ell,\tilde\ell') \int_{\gamma(\tilde\ell,\tilde\ell')} d\tau\,V^{(s_4,s_5)}_{\text{new}}\big(\del_{x_4},\del_{u_4};\del_{x_5},\del_{u_5};\dot x(\tau)\big) \\
        &\qquad\qquad\qquad\qquad \times \Pi^{(s_4)}(x_4,u_4;\ell_4,\lambda_4)\,\Pi^{(s_5)}(x_5,u_5;\ell_5,\lambda_5) \Big|_{x_4=x_5=x(\tau)} \ .
    \end{split}
\end{align}
Diagram no. 6: $\vcenter{\includegraphics[scale=0.2]{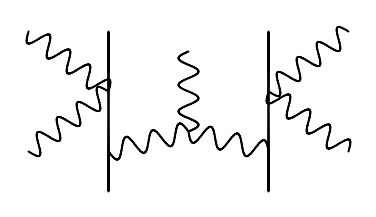}}$
It has $5\cdot\binom{4}{2}/2 = 15$ inequivalent orderings of the external legs. The expression for one sample ordering reads:
\begin{align}
    \begin{split}
        &\left(\frac{N}{4}\right)^2 \int d^3\ell\,d^3\ell'\,G(\ell,\ell') \int_{\gamma(\ell,\ell')}  d\tau_{\text{left}}\,V^{(s_1,s_2)}_{\text{new}}\big(\del_{x_1},\del_{u_1};\del_{x_2},\del_{u_2};\dot x(\tau_{\text{left}})\big) \\
        &\qquad\qquad\qquad\qquad \times \Pi^{(s_1)}(x_1,u_1;\ell_1,\lambda_1)\,\Pi^{(s_2)}(x_2,u_2;\ell_2,\lambda_2) \Big|_{x_1=x_2=x(\tau_{\text{left}})} \\
        &\quad \times \Box_\ell\Box_{\ell'}\,G(\ell,\ell')\int d^3\tilde\ell\,d^3\tilde\ell'\,G(\tilde\ell,\tilde\ell')\sum_{s_6,s_7}\int_{EAdS_4}
        d^4x\,V^{(s_3,s_6,s_7)}(\del_{x_3},\del_{u_3};\del_{x_6},\del_{u_6};\del_{x_7},\del_{u_7}) \\
        &\qquad\qquad\qquad\qquad \times \Pi^{(s_3)}(x_3,u_3;\ell_3,\lambda_3)\,\phi^{(s_6)}(x_6,u_6;\ell,\ell')\,\phi^{(s_7)}(x_7,u_7;\tilde\ell,\tilde\ell') \Big|_{x_3=x_6=x_7=x} \\
        &\quad \times \Box_{\tilde\ell}\Box_{\tilde\ell'}\,G(\tilde\ell,\tilde\ell') \int_{\gamma(\tilde\ell,\tilde\ell')} d\tau_{\text{right}}\,
        V^{(s_4,s_5)}_{\text{new}}\big(\del_{x_4},\del_{u_4};\del_{x_5},\del_{u_5};\dot x(\tau_{\text{right}})\big) \\
        &\qquad\qquad\qquad\qquad \times \Pi^{(s_4)}(x_4,u_4;\ell_4,\lambda_4)\,\Pi^{(s_5)}(x_5,u_5;\ell_5,\lambda_5) \Big|_{x_4=x_5=x(\tau_{\text{right}})} \ .
    \end{split}
\end{align}
Diagram no. 7: $\vcenter{\includegraphics[scale=0.2]{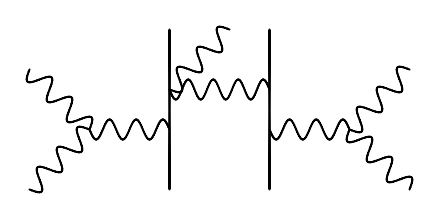}}$
It has $5\cdot\binom{4}{2} = 30$ inequivalent orderings of the external legs. The expression for one sample ordering reads:
\begin{align}
    \begin{split}
        &\left(\frac{N}{4}\right)^2 \int d^3\ell\,d^3\ell'\,G(\ell,\ell')\sum_{s_6}\int_{EAdS_4} d^4x_{\text{left}}\,V^{(s_1,s_2,s_6)}(\del_{x_1},\del_{u_1};\del_{x_2},\del_{u_2};\del_{x_6},\del_{u_6}) \\
        &\qquad\qquad\qquad\qquad \times \Pi^{(s_1)}(x_1,u_1;\ell_1,\lambda_1)\,\Pi^{(s_2)}(x_2,u_2;\ell_2,\lambda_2)\,\phi^{(s_6)}(x_6,u_6;\ell,\ell') \Big|_{x_1=x_2=x_6=x_{\text{left}}} \\
        &\quad \times \Box_\ell\Box_{\ell'}\,G(\ell,\ell')\int d^3\tilde\ell\,d^3\tilde\ell'\,G(\tilde\ell,\tilde\ell') \sum_{s_7} 
        \int_{\gamma(\ell,\ell')} d\tau\,V^{(s_3,s_7)}_{\text{new}}\big(\del_{x_3},\del_{u_3};\del_{x_7},\del_{u_7};\dot x(\tau)\big) \\
        &\qquad\qquad\qquad\qquad \times \Pi^{(s_3)}\big(x_3,u_3;\ell_3,\lambda_3)\,\phi^{(s_7)}\big(x_7,u_7;\tilde\ell,\tilde\ell') \Big|_{x_3=x_7=x(\tau)}  \\
        &\quad \times \Box_{\tilde\ell}\Box_{\tilde\ell'}\,G(\tilde\ell,\tilde\ell') \sum_{s_8}\int_{EAdS_4} d^4x_{\text{right}}\,V^{(s_4,s_5,s_8)}(\del_{x_4},\del_{u_4};\del_{x_5},\del_{u_5};\del_{x_8},\del_{u_8}) \\
        &\qquad\qquad\qquad\qquad \times \Pi^{(s_4)}(x_4,u_4;\ell_4,\lambda_4)\,\Pi^{(s_5)}(x_5,u_5;\ell_5,\lambda_5)\,\phi^{(s_8)}(x_8,u_8;\tilde\ell,\tilde\ell') \Big|_{x_4=x_5=x_8=x_{\text{right}}} \ .
    \end{split}
\end{align}
Diagram no. 8: $\vcenter{\includegraphics[scale=0.2]{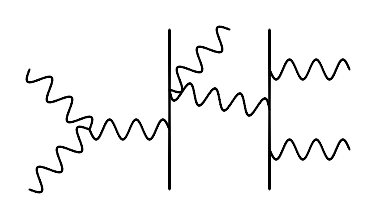}}$
It has $5\cdot\binom{4}{2} = 30$ inequivalent orderings of the external legs. The expression for one sample ordering reads:
\begin{align}
    \begin{split}
        &\left(\frac{N}{4}\right)^2 \int d^3\ell\,d^3\ell'\,G(\ell,\ell')\sum_{s_6}\int_{EAdS_4} d^4x\,V^{(s_1,s_2,s_6)}(\del_{x_1},\del_{u_1};\del_{x_2},\del_{u_2};\del_{x_6},\del_{u_6}) \\
        &\qquad\qquad\qquad\qquad \times \Pi^{(s_1)}(x_1,u_1;\ell_1,\lambda_1)\,\Pi^{(s_2)}(x_2,u_2;\ell_2,\lambda_2)\,\phi^{(s_6)}(x_6,u_6;\ell,\ell') \Big|_{x_1=x_2=x_6=x} \\
        &\quad \times \Box_\ell\Box_{\ell'}\,G(\ell,\ell')\int d^3\tilde\ell\,d^3\tilde\ell'\,G(\tilde\ell,\tilde\ell') \sum_{s_7} 
        \int_{\gamma(\ell,\ell')} d\tau\,V^{(s_3,s_7)}_{\text{new}}\big(\del_{x_3},\del_{u_3};\del_{x_7},\del_{u_7};\dot x(\tau)\big) \\
        &\qquad\qquad\qquad\qquad \times \Pi^{(s_3)}\big(x_3,u_3;\ell_3,\lambda_3)\,\phi^{(s_7)}\big(x_7,u_7;\tilde\ell,\tilde\ell') \Big|_{x_3=x_7=x(\tau)}  \\
        &\quad \times \Box_{\tilde\ell}\Box_{\tilde\ell'}\,G(\tilde\ell,\tilde\ell')\,Q^{(s_4)}\int_{\gamma(\tilde\ell,\tilde\ell')} d\tau_4\,\big(\dot x(\tau_4)\cdot\del_{u_4}\big)^{s_4}\,
        \Pi^{(s_4)}\big(x(\tau_4),u_4;\ell_4,\lambda_4\big) \\
        &\qquad\qquad \times Q^{(s_5)}\int_{\gamma(\tilde\ell,\tilde\ell')} d\tau_5\,\big(\dot x(\tau_5)\cdot\del_{u_5}\big)^{s_5}\,\Pi^{(s_5)}\big(x(\tau_5),u_5;\ell_5,\lambda_5\big) \ .
    \end{split}
\end{align}
Diagram no. 9: $\vcenter{\includegraphics[scale=0.2]{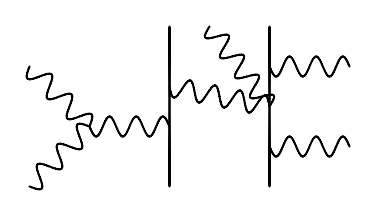}}$
It has $5\cdot\binom{4}{2} = 30$ inequivalent orderings of the external legs. The expression for one sample ordering reads:
\begin{align}
    \begin{split}
        &\left(\frac{N}{4}\right)^2 \int d^3\ell\,d^3\ell'\,G(\ell,\ell')\sum_{s_6}\int_{EAdS_4} d^4x\,V^{(s_1,s_2,s_6)}(\del_{x_1},\del_{u_1};\del_{x_2},\del_{u_2};\del_{x_6},\del_{u_6}) \\
        &\qquad\qquad\qquad\qquad \times \Pi^{(s_1)}(x_1,u_1;\ell_1,\lambda_1)\,\Pi^{(s_2)}(x_2,u_2;\ell_2,\lambda_2)\,\phi^{(s_6)}(x_6,u_6;\ell,\ell') \Big|_{x_1=x_2=x_6=x} \\
        &\quad \times \Box_\ell\Box_{\ell'}\,G(\ell,\ell')\int d^3\tilde\ell\,d^3\tilde\ell'\,G(\tilde\ell,\tilde\ell') \sum_{s_7} 
        \int_{\gamma(\tilde\ell,\tilde\ell')} d\tau\,V^{(s_3,s_7)}_{\text{new}}\big(\del_{x_3},\del_{u_3};\del_{x_7},\del_{u_7};\dot x(\tau)\big) \\
        &\qquad\qquad\qquad\qquad \times \Pi^{(s_3)}\big(x_3,u_3;\ell_3,\lambda_3)\,\phi^{(s_7)}\big(x_7,u_7;\ell,\ell') \Big|_{x_3=x_7=x(\tau)}  \\
        &\quad \times \Box_{\tilde\ell}\Box_{\tilde\ell'}\,G(\tilde\ell,\tilde\ell')\,Q^{(s_4)}\int_{\gamma(\tilde\ell,\tilde\ell')} d\tau_4\,\big(\dot x(\tau_4)\cdot\del_{u_4}\big)^{s_4}\,
        \Pi^{(s_4)}\big(x(\tau_4),u_4;\ell_4,\lambda_4\big) \\
        &\qquad\qquad \times Q^{(s_5)}\int_{\gamma(\tilde\ell,\tilde\ell')} d\tau_5\,\big(\dot x(\tau_5)\cdot\del_{u_5}\big)^{s_5}\,\Pi^{(s_5)}\big(x(\tau_5),u_5;\ell_5,\lambda_5\big) \ .
    \end{split}
\end{align}
Diagram no. 10: $\vcenter{\includegraphics[scale=0.2]{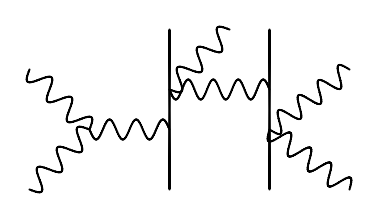}}$
It has $5\cdot\binom{4}{2} = 30$ inequivalent orderings of the external legs. The expression for one sample ordering reads:
\begin{align}
    \begin{split}
        &\left(\frac{N}{4}\right)^2 \int d^3\ell\,d^3\ell'\,G(\ell,\ell')\sum_{s_6}\int_{EAdS_4} d^4x\,V^{(s_1,s_2,s_6)}(\del_{x_1},\del_{u_1};\del_{x_2},\del_{u_2};\del_{x_6},\del_{u_6}) \\
        &\qquad\qquad\qquad\qquad \times \Pi^{(s_1)}(x_1,u_1;\ell_1,\lambda_1)\,\Pi^{(s_2)}(x_2,u_2;\ell_2,\lambda_2)\,\phi^{(s_6)}(x_6,u_6;\ell,\ell') \Big|_{x_1=x_2=x_6=x} \\
        &\quad \times \Box_\ell\Box_{\ell'}\,G(\ell,\ell')\int d^3\tilde\ell\,d^3\tilde\ell'\,G(\tilde\ell,\tilde\ell') \sum_{s_7} 
        \int_{\gamma(\ell,\ell')} d\tau_{\text{mid}}\,V^{(s_3,s_7)}_{\text{new}}\big(\del_{x_3},\del_{u_3};\del_{x_7},\del_{u_7};\dot x(\tau_{\text{mid}})\big) \\
        &\qquad\qquad\qquad\qquad \times \Pi^{(s_3)}(x_3,u_3;\ell_3,\lambda_3)\,\phi^{(s_7)}\big(x_7,u_7;\tilde\ell,\tilde\ell') \Big|_{x_3=x_7=x(\tau_{\text{mid}})}  \\
        &\quad \times \Box_{\tilde\ell}\Box_{\tilde\ell'}\,G(\tilde\ell,\tilde\ell') \int_{\gamma(\tilde\ell,\tilde\ell')} d\tau_{\text{right}}\,
        V^{(s_4,s_5)}_{\text{new}}\big(\del_{x_4},\del_{u_4};\del_{x_5},\del_{u_5};\dot x(\tau_{\text{right}})\big) \\
        &\qquad\qquad\qquad\qquad \times \Pi^{(s_4)}(x_4,u_4;\ell_4,\lambda_4)\,\Pi^{(s_5)}\big(x_5,u_5;\ell_5,\lambda_5) \Big|_{x_4=x_5=x(\tau_{\text{right}})} \ .
    \end{split}
\end{align}
Diagram no. 11: $\vcenter{\includegraphics[scale=0.2]{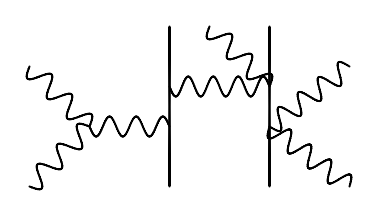}}$
It has $5\cdot\binom{4}{2} = 30$ inequivalent orderings of the external legs. The expression for one sample ordering reads:
\begin{align}
    \begin{split}
        &\left(\frac{N}{4}\right)^2 \int d^3\ell\,d^3\ell'\,G(\ell,\ell')\sum_{s_6}\int_{EAdS_4} d^4x\,V^{(s_1,s_2,s_6)}(\del_{x_1},\del_{u_1};\del_{x_2},\del_{u_2};\del_{x_6},\del_{u_6}) \\
        &\qquad\qquad\qquad\qquad \times \Pi^{(s_1)}(x_1,u_1;\ell_1,\lambda_1)\,\Pi^{(s_2)}(x_2,u_2;\ell_2,\lambda_2)\,\phi^{(s_6)}(x_6,u_6;\ell,\ell') \Big|_{x_1=x_2=x_6=x} \\
        &\quad \times \Box_\ell\Box_{\ell'}\,G(\ell,\ell')\int d^3\tilde\ell\,d^3\tilde\ell'\,G(\tilde\ell,\tilde\ell') \sum_{s_7} 
        \int_{\gamma(\tilde\ell,\tilde\ell')} d\tau_{\text{mid}}\,V^{(s_3,s_7)}_{\text{new}}\big(\del_{x_3},\del_{u_3};\del_{x_7},\del_{u_7};\dot x(\tau_{\text{mid}})\big) \\
        &\qquad\qquad \times \Pi^{(s_3)}(x_3,u_3;\ell_3,\lambda_3)\,\phi^{(s_7)}\big(x_7,u_7;\ell,\ell') \Big|_{x_3=x_7=x(\tau_{\text{mid}})}  \\
        &\quad \times \Box_{\tilde\ell}\Box_{\tilde\ell'}\,G(\tilde\ell,\tilde\ell') \int_{\gamma(\tilde\ell,\tilde\ell')} d\tau_{\text{right}}\,
        V^{(s_4,s_5)}_{\text{new}}\big(\del_{x_4},\del_{u_4};\del_{x_5},\del_{u_5};\dot x(\tau_{\text{right}})\big) \\
        &\qquad\qquad\qquad\qquad \times \Pi^{(s_4)}(x_4,u_4;\ell_4,\lambda_4)\,\Pi^{(s_5)}\big(x_5,u_5;\ell_5,\lambda_5) \Big|_{x_4=x_5=x(\tau_{\text{right}})} \ .
    \end{split}
\end{align}
Diagram no. 12: $\vcenter{\includegraphics[scale=0.2]{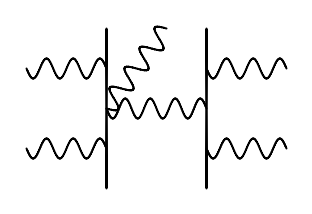}}$
It has $5\cdot\binom{4}{2} = 30$ inequivalent orderings of the external legs. The expression for one sample ordering reads:
\begin{align}
    \begin{split}
        &\left(\frac{N}{4}\right)^2 \int d^3\ell\,d^3\ell'\,G(\ell,\ell')\,Q^{(s_1)}\int_{\gamma(\ell,\ell')} d\tau_1\,\big(\dot x(\tau_1)\cdot\del_{u_1}\big)^{s_1}\,\Pi^{(s_1)}\big(x(\tau_1),u_1;\ell_1,\lambda_1\big) \\
        &\qquad\qquad \times Q^{(s_2)}\int_{\gamma(\ell,\ell')} d\tau_2\,\big(\dot x(\tau_2)\cdot\del_{u_2}\big)^{s_2}\,\Pi^{(s_2)}\big(x(\tau_2),u_2;\ell_2,\lambda_2\big) \\
        &\quad \times \Box_\ell\Box_{\ell'}\,G(\ell,\ell')\int d^3\tilde\ell\,d^3\tilde\ell'\,G(\tilde\ell,\tilde\ell') \sum_{s_6} 
        \int_{\gamma(\ell,\ell')} d\tau\,V^{(s_3,s_6)}_{\text{new}}\big(\del_{x_3},\del_{u_3};\del_{x_6},\del_{u_6};\dot x(\tau)\big) \\
        &\qquad\qquad\qquad\qquad \times \Pi^{(s_3)}(x_3,u_3;\ell_3,\lambda_3)\,\phi^{(s_6)}\big(x_6,u_6;\tilde\ell,\tilde\ell') \Big|_{x_3=x_6=x(\tau)}  \\
        &\quad \times \Box_{\tilde\ell}\Box_{\tilde\ell'}\,G(\tilde\ell,\tilde\ell')\,Q^{(s_4)}\int_{\gamma(\tilde\ell,\tilde\ell')} d\tau_4\,\big(\dot x(\tau_4)\cdot\del_{u_4}\big)^{s_4}\,
        \Pi^{(s_4)}\big(x(\tau_4),u_4;\ell_4,\lambda_4\big) \\
        &\qquad\qquad \times Q^{(s_5)}\int_{\gamma(\tilde\ell,\tilde\ell')} d\tau_5\,\big(\dot x(\tau_5)\cdot\del_{u_5}\big)^{s_5}\,\Pi^{(s_5)}\big(x(\tau_5),u_5;\ell_5,\lambda_5\big) \ .
    \end{split}
\end{align}
Diagram no. 13: $\vcenter{\includegraphics[scale=0.2]{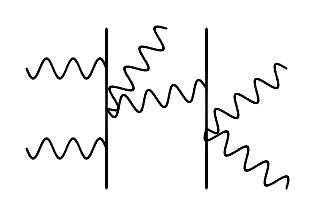}}$
It has $5\cdot\binom{4}{2} = 30$ inequivalent orderings of the external legs. The expression for one sample ordering reads:
\begin{align}
    \begin{split}
        &\left(\frac{N}{4}\right)^2 \int d^3\ell\,d^3\ell'\,G(\ell,\ell')\,Q^{(s_1)}\int_{\gamma(\ell,\ell')} d\tau_1\,\big(\dot x(\tau_1)\cdot\del_{u_1}\big)^{s_1}\,\Pi^{(s_1)}\big(x(\tau_1),u_1;\ell_1,\lambda_1\big) \\
        &\qquad\qquad \times Q^{(s_2)}\int_{\gamma(\ell,\ell')} d\tau_2\,\big(\dot x(\tau_2)\cdot\del_{u_2}\big)^{s_2}\,\Pi^{(s_2)}\big(x(\tau_2),u_2;\ell_2,\lambda_2\big) \\
        &\quad \times \Box_\ell\Box_{\ell'}\,G(\ell,\ell')\int d^3\tilde\ell\,d^3\tilde\ell'\,G(\tilde\ell,\tilde\ell') \sum_{s_6} 
        \int_{\gamma(\ell,\ell')} d\tau_{\text{mid}}\,V^{(s_3,s_6)}_{\text{new}}\big(\del_{x_3},\del_{u_3};\del_{x_6},\del_{u_6};\dot x(\tau_{\text{mid}})\big) \\
        &\qquad\qquad\qquad\qquad \times \Pi^{(s_3)}(x_3,u_3;\ell_3,\lambda_3)\,\phi^{(s_6)}\big(x_6,u_6;\tilde\ell,\tilde\ell') \Big|_{x_3=x_6=x(\tau_{\text{mid}})}  \\
        &\quad \times \Box_{\tilde\ell}\Box_{\tilde\ell'}\,G(\tilde\ell,\tilde\ell') \int_{\gamma(\tilde\ell,\tilde\ell')} d\tau_{\text{right}}\,
        V^{(s_4,s_5)}_{\text{new}}\big(\del_{x_4},\del_{u_4};\del_{x_5},\del_{u_5};\dot x(\tau_{\text{right}})\big) \\
        &\qquad\qquad\qquad\qquad \times \Pi^{(s_4)}(x_4,u_4;\ell_4,\lambda_4)\,\Pi^{(s_5)}(x_5,u_5;\ell_5,\lambda_5) \Big|_{x_4=x_5=x(\tau_{\text{right}})} \ .
    \end{split}
\end{align}
Diagram no. 14: $\vcenter{\includegraphics[scale=0.2]{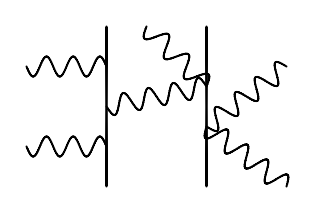}}$
It has $5\cdot\binom{4}{2} = 30$ inequivalent orderings of the external legs. The expression for one sample ordering reads:
\begin{align}
    \begin{split}
        &\left(\frac{N}{4}\right)^2 \int d^3\ell\,d^3\ell'\,G(\ell,\ell')\,Q^{(s_1)}\int_{\gamma(\ell,\ell')} d\tau_1\,\big(\dot x(\tau_1)\cdot\del_{u_1}\big)^{s_1}\,\Pi^{(s_1)}\big(x(\tau_1),u_1;\ell_1,\lambda_1\big) \\
        &\qquad\qquad \times Q^{(s_2)}\int_{\gamma(\ell,\ell')} d\tau_2\,\big(\dot x(\tau_2)\cdot\del_{u_2}\big)^{s_2}\,\Pi^{(s_2)}\big(x(\tau_2),u_2;\ell_2,\lambda_2\big) \\
        &\quad \times \Box_\ell\Box_{\ell'}\,G(\ell,\ell')\int d^3\tilde\ell\,d^3\tilde\ell'\,G(\tilde\ell,\tilde\ell') \sum_{s_6} 
        \int_{\gamma(\tilde\ell,\tilde\ell')} d\tau_{\text{mid}}\,V^{(s_3,s_6)}_{\text{new}}\big(\del_{x_3},\del_{u_3};\del_{x_6},\del_{u_6};\dot x(\tau_{\text{mid}})\big) \\
        &\qquad\qquad\qquad\qquad \times \Pi^{(s_3)}(x_3,u_3;\ell_3,\lambda_3)\,\phi^{(s_6)}\big(x_6,u_6;\ell,\ell') \Big|_{x_3=x_6=x(\tau_{\text{mid}})}  \\
        &\quad \times \Box_{\tilde\ell}\Box_{\tilde\ell'}\,G(\tilde\ell,\tilde\ell') \int_{\gamma(\tilde\ell,\tilde\ell')} d\tau_{\text{right}}\,
        V^{(s_4,s_5)}_{\text{new}}\big(\del_{x_4},\del_{u_4};\del_{x_5},\del_{u_5};\dot x(\tau_{\text{right}})\big) \\
        &\qquad\qquad\qquad\qquad \times \Pi^{(s_4)}(x_4,u_4;\ell_4,\lambda_4)\,\Pi^{(s_5)}(x_5,u_5;\ell_5,\lambda_5) \Big|_{x_4=x_5=x(\tau_{\text{right}})} \ .
    \end{split}
\end{align}
Diagram no. 15: $\vcenter{\includegraphics[scale=0.2]{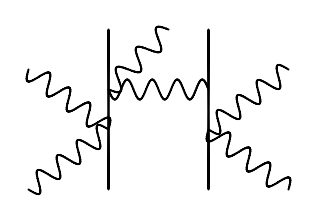}}$
It has $5\cdot\binom{4}{2} = 30$ inequivalent orderings of the external legs. The expression for one sample ordering reads:
\begin{align}
    \begin{split}
        &\left(\frac{N}{4}\right)^2 \int d^3\ell\,d^3\ell'\,G(\ell,\ell') \int_{\gamma(\ell,\ell')} d\tau_{\text{left}}\,
        V^{(s_1,s_2)}_{\text{new}}\big(\del_{x_1},\del_{u_1};\del_{x_2},\del_{u_2};\dot x(\tau_{\text{left}})\big) \\
        &\qquad\qquad\qquad\qquad \times \Pi^{(s_1)}(x_1,u_1;\ell_1,\lambda_1)\,\Pi^{(s_2)}(x_2,u_2;\ell_2,\lambda_2) \Big|_{x_1=x_2=x(\tau_{\text{left}})} \\
        &\quad \times \Box_\ell\Box_{\ell'}\,G(\ell,\ell')\int d^3\tilde\ell\,d^3\tilde\ell'\,G(\tilde\ell,\tilde\ell') \sum_{s_6} 
        \int_{\gamma(\ell,\ell')} d\tau_{\text{mid}}\,V^{(s_3,s_6)}_{\text{new}}\big(\del_{x_3},\del_{u_3};\del_{x_6},\del_{u_6};\dot x(\tau_{\text{mid}})\big) \\
        &\qquad\qquad\qquad\qquad \times \Pi^{(s_3)}(x_3,u_3;\ell_3,\lambda_3)\,\phi^{(s_6)}\big(x_6,u_6;\tilde\ell,\tilde\ell') \Big|_{x_3=x_6=x(\tau_{\text{mid}})}  \\
        &\quad \times \Box_{\tilde\ell}\Box_{\tilde\ell'}\,G(\tilde\ell,\tilde\ell') \int_{\gamma(\tilde\ell,\tilde\ell')} d\tau_{\text{right}}\,
        V^{(s_4,s_5)}_{\text{new}}\big(\del_{x_4},\del_{u_4};\del_{x_5},\del_{u_5};\dot x(\tau_{\text{right}})\big) \\
        &\qquad\qquad\qquad\qquad \times \Pi^{(s_4)}(x_4,u_4;\ell_4,\lambda_4)\,\Pi^{(s_5)}(x_5,u_5;\ell_5,\lambda_5) \Big|_{x_4=x_5=x(\tau_{\text{right}})} \ .
    \end{split}
\end{align}
Diagram no. 16: $\vcenter{\includegraphics[scale=0.2]{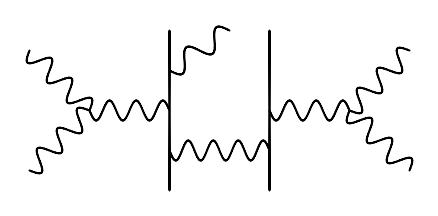}}$
After identifying the worldlines, this contracts into: $\vcenter{\includegraphics[scale=0.2]{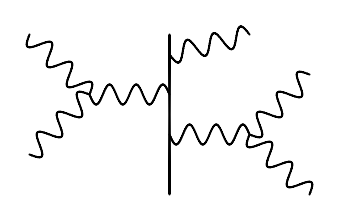}}$
Both diagrams have $5\cdot\binom{4}{2} = 30$ inequivalent orderings of the external legs. For one sample ordering, the contracted diagram reads:
\begin{align}
    \begin{split}
        &\frac{1}{2}\times\frac{N}{4} \int d^3\ell\,d^3\ell'\,G(\ell,\ell')\sum_{s_6}\int_{EAdS_4} d^4x_{\text{left}}\,V^{(s_1,s_2,s_6)}(\del_{x_1},\del_{u_1};\del_{x_2},\del_{u_2};\del_{x_6},\del_{u_6}) \\
        &\qquad\qquad\qquad\qquad \times \Pi^{(s_1)}(x_1,u_1;\ell_1,\lambda_1)\,\Pi^{(s_2)}(x_2,u_2;\ell_2,\lambda_2)\,\phi^{(s_6)}(x_6,u_6;\ell,\ell') \Big|_{x_1=x_2=x_6=x_{\text{left}}} \\
        &\quad \times \Box_\ell\Box_{\ell'}\,G(\ell,\ell')\,Q^{(s_3)}\int_{\gamma(\ell,\ell')} d\tau\,\big(\dot x(\tau)\cdot\del_{u}\big)^{s_3}\,\Pi^{(s_3)}\big(x(\tau),u;\ell_3,\lambda_3\big) \\
        &\quad \times \sum_{s_7}\int_{EAdS_4} d^4x_{\text{right}}\,V^{(s_4,s_5,s_7)}(\del_{x_4},\del_{u_4};\del_{x_5},\del_{u_5};\del_{x_7},\del_{u_7}) \\
        &\qquad\qquad\qquad\qquad \times \Pi^{(s_4)}(x_4,u_4;\ell_4,\lambda_4)\,\Pi^{(s_5)}(x_5,u_5;\ell_5,\lambda_5)\,\phi^{(s_7)}(x_7,u_7;\ell,\ell') \Big|_{x_4=x_5=x_7=x_{\text{right}}} \ .
    \end{split}
\end{align}
Diagram no. 17: $\vcenter{\includegraphics[scale=0.2]{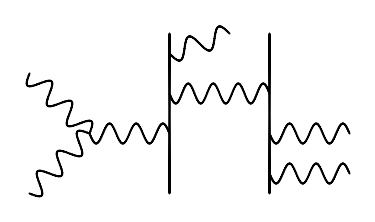}}$
After identifying the worldlines, this contracts into: $\vcenter{\includegraphics[scale=0.2]{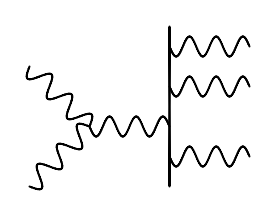}}$
The original diagram has $5\cdot\binom{4}{2} = 30$ inequivalent orderings of the external legs, but the contracted diagram has only $\binom{5}{2} = 10$. Thus, when we sum over inequivalent orderings in the contracted diagram, we should multiply by a factor of $3$.  For one sample ordering, the contracted diagram reads:
\begin{align}
    \begin{split}
        &\frac{3}{2}\times\frac{N}{4} \int d^3\ell\,d^3\ell'\,G(\ell,\ell')\sum_{s_6}\int_{EAdS_4} d^4x\,V^{(s_1,s_2,s_6)}(\del_{x_1},\del_{u_1};\del_{x_2},\del_{u_2};\del_{x_6},\del_{u_6}) \\
        &\qquad\qquad\qquad\qquad \times \Pi^{(s_1)}(x_1,u_1;\ell_1,\lambda_1)\,\Pi^{(s_2)}(x_2,u_2;\ell_2,\lambda_2)\,\phi^{(s_6)}(x_6,u_6;\ell,\ell') \Big|_{x_1=x_2=x_6=x} \\
        &\quad \times \Box_\ell\Box_{\ell'}\,G(\ell,\ell')\,Q^{(s_3)}\int_{\gamma(\ell,\ell')} d\tau_3\,\big(\dot x(\tau_3)\cdot\del_{u_3}\big)^{s_3}\,\Pi^{(s_3)}\big(x(\tau_3),u_3;\ell_3,\lambda_3\big) \\
        &\qquad\qquad \times Q^{(s_4)}\int_{\gamma(\ell,\ell')} d\tau_4\,\big(\dot x(\tau_4)\cdot\del_{u_4}\big)^{s_4}\,\Pi^{(s_4)}\big(x(\tau_4),u_4;\ell_4,\lambda_4\big) \\
        &\qquad\qquad \times Q^{(s_5)}\int_{\gamma(\ell,\ell')} d\tau_5\,\big(\dot x(\tau_5)\cdot\del_{u_5}\big)^{s_5}\,\Pi^{(s_5)}\big(x(\tau_5),u_5;\ell_5,\lambda_5\big) \ .
    \end{split}
\end{align}
Diagram no. 18: $\vcenter{\includegraphics[scale=0.2]{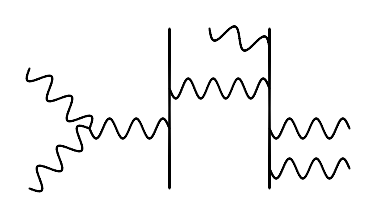}}$
After identifying the worldlines, this contracts into: $\vcenter{\includegraphics[scale=0.2]{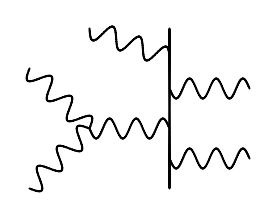}}$
Both diagrams have $5\cdot\binom{4}{2} = 30$ inequivalent orderings of the external legs. For one sample ordering, the contracted diagram reads:
\begin{align}
    \begin{split}
        &\frac{1}{2}\times\frac{N}{4} \int d^3\ell\,d^3\ell'\,G(\ell,\ell')\sum_{s_6}\int_{EAdS_4} d^4x\,V^{(s_1,s_2,s_6)}(\del_{x_1},\del_{u_1};\del_{x_2},\del_{u_2};\del_{x_6},\del_{u_6}) \\
        &\qquad\qquad\qquad\qquad \times \Pi^{(s_1)}(x_1,u_1;\ell_1,\lambda_1)\,\Pi^{(s_2)}(x_2,u_2;\ell_2,\lambda_2)\,\phi^{(s_6)}(x_6,u_6;\ell,\ell') \Big|_{x_1=x_2=x_6=x} \\
        &\quad \times Q^{(s_3)}\int_{\gamma(\ell,\ell')} d\tau_3\,\big(\dot x(\tau_3)\cdot\del_{u_3}\big)^{s_3}\,\Pi^{(s_3)}\big(x(\tau_3),u_3;\ell_3,\lambda_3\big) \\
        &\quad \times \Box_\ell\Box_{\ell'}\,G(\ell,\ell')\,Q^{(s_4)}\int_{\gamma(\ell,\ell')} d\tau_4\,\big(\dot x(\tau_4)\cdot\del_{u_4}\big)^{s_4}\,\Pi^{(s_4)}\big(x(\tau_4),u_4;\ell_4,\lambda_4\big) \\
        &\qquad\qquad \times Q^{(s_5)}\int_{\gamma(\ell,\ell')} d\tau_5\,\big(\dot x(\tau_5)\cdot\del_{u_5}\big)^{s_5}\,\Pi^{(s_5)}\big(x(\tau_5),u_5;\ell_5,\lambda_5\big) \ .
    \end{split}
\end{align}
Diagram no. 19: $\vcenter{\includegraphics[scale=0.2]{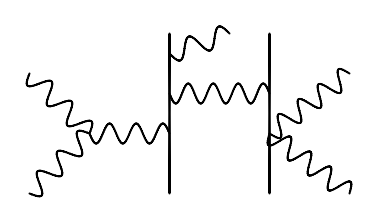}}$
After identifying the worldlines, this contracts into: $\vcenter{\includegraphics[scale=0.2]{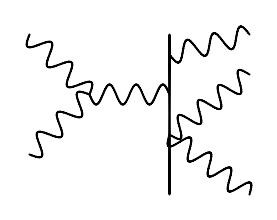}}$
Both diagrams have $5\cdot\binom{4}{2} = 30$ inequivalent orderings of the external legs. For one sample ordering, the contracted diagram reads:
\begin{align}
    \begin{split}
        &\frac{1}{2}\times\frac{N}{4} \int d^3\ell\,d^3\ell'\,G(\ell,\ell')\sum_{s_6}\int_{EAdS_4} d^4x\,V^{(s_1,s_2,s_6)}(\del_{x_1},\del_{u_1};\del_{x_2},\del_{u_2};\del_{x_6},\del_{u_6}) \\
        &\qquad\qquad\qquad\qquad \times \Pi^{(s_1)}(x_1,u_1;\ell_1,\lambda_1)\,\Pi^{(s_2)}(x_2,u_2;\ell_2,\lambda_2)\,\phi^{(s_6)}(x_6,u_6;\ell,\ell') \Big|_{x_1=x_2=x_6=x} \\
        &\quad \times \Box_\ell\Box_{\ell'}\,G(\ell,\ell')\,Q^{(s_3)}\int_{\gamma(\ell,\ell')} d\tau\,\big(\dot x(\tau)\cdot\del_{u}\big)^{s_3}\,\Pi^{(s_3)}\big(x(\tau),u;\ell_3,\lambda_3\big) \\
        &\quad \times \int_{\gamma(\ell,\ell')} d\tau\,V^{(s_4,s_5)}_{\text{new}}\big(\del_{x_4},\del_{u_4};\del_{x_5},\del_{u_5};\dot x(\tau)\big) \\
        &\qquad\qquad\qquad\qquad \times \Pi^{(s_4)}(x_4,u_4;\ell_4,\lambda_4)\,\Pi^{(s_5)}(x_5,u_5;\ell_5,\lambda_5) \Big|_{x_4=x_5=x(\tau)} \ .
    \end{split}
\end{align}
Diagram no. 20: $\vcenter{\includegraphics[scale=0.2]{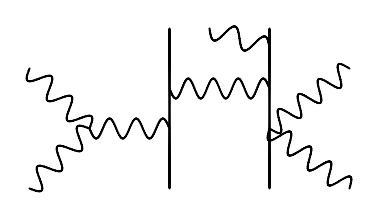}}$
After identifying the worldlines, this contracts into: $\vcenter{\includegraphics[scale=0.2]{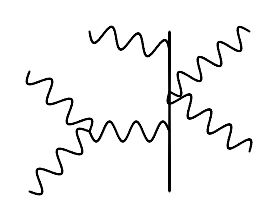}}$
Both diagrams have $5\cdot\binom{4}{2} = 30$ inequivalent orderings of the external legs. For one sample ordering, the contracted diagram reads:
\begin{align}
    \begin{split}
        &\frac{1}{2}\times\frac{N}{4} \int d^3\ell\,d^3\ell'\,G(\ell,\ell')\sum_{s_6}\int_{EAdS_4} d^4x\,V^{(s_1,s_2,s_6)}(\del_{x_1},\del_{u_1};\del_{x_2},\del_{u_2};\del_{x_6},\del_{u_6}) \\
        &\qquad\qquad\qquad\qquad \times \Pi^{(s_1)}(x_1,u_1;\ell_1,\lambda_1)\,\Pi^{(s_2)}(x_2,u_2;\ell_2,\lambda_2)\,\phi^{(s_6)}(x_6,u_6;\ell,\ell') \Big|_{x_1=x_2=x_6=x} \\
        &\quad \times Q^{(s_3)}\int_{\gamma(\ell,\ell')} d\tau\,\big(\dot x(\tau)\cdot\del_{u}\big)^{s_3}\,\Pi^{(s_3)}\big(x(\tau),u;\ell_3,\lambda_3\big) \\
        &\quad \times \Box_\ell\Box_{\ell'}\,G(\ell,\ell') \int_{\gamma(\ell,\ell')} d\tau\,V^{(s_4,s_5)}_{\text{new}}\big(\del_{x_4},\del_{u_4};\del_{x_5},\del_{u_5};\dot x(\tau)\big) \\
        &\qquad\qquad\qquad\qquad \times \Pi^{(s_4)}(x_4,u_4;\ell_4,\lambda_4)\,\Pi^{(s_5)}(x_5,u_5;\ell_5,\lambda_5) \Big|_{x_4=x_5=x(\tau)} \ .
    \end{split}
\end{align}
Diagram no. 21: $\vcenter{\includegraphics[scale=0.2]{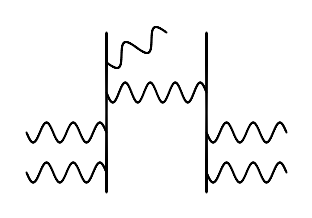}}$
After identifying the worldlines, this contracts into: $\vcenter{\includegraphics[scale=0.2]{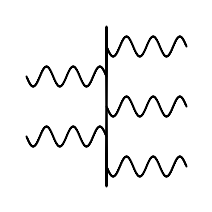}}$
The original diagram has $5\cdot\binom{4}{2} = 30$ inequivalent orderings of the external legs, but the contracted diagram has only $\binom{5}{2} = 10$. Thus, when we sum over inequivalent orderings in the contracted diagram, we should multiply by a factor of $3$.  For one sample ordering, the contracted diagram reads:
\begin{align}
    \begin{split}
        &\frac{3}{2}\times\frac{N}{4} \int d^3\ell\,d^3\ell'\,G(\ell,\ell')\,Q^{(s_1)}\int_{\gamma(\ell,\ell')} d\tau_1\,\big(\dot x(\tau_1)\cdot\del_{u_1}\big)^{s_1}\,\Pi^{(s_1)}\big(x(\tau_1),u_1;\ell_1,\lambda_1\big) \\
        &\qquad\qquad \times Q^{(s_2)}\int_{\gamma(\ell,\ell')} d\tau_2\,\big(\dot x(\tau_2)\cdot\del_{u_2}\big)^{s_2}\,\Pi^{(s_2)}\big(x(\tau_2),u_2;\ell_2,\lambda_2\big) \\
        &\quad \times \Box_\ell\Box_{\ell'}\,G(\ell,\ell')\,Q^{(s_3)}\int_{\gamma(\ell,\ell')} d\tau_3\,\big(\dot x(\tau_3)\cdot\del_{u_3}\big)^{s_3}\,\Pi^{(s_3)}\big(x(\tau_3),u_3;\ell_3,\lambda_3\big) \\
        &\qquad\qquad \times Q^{(s_4)}\int_{\gamma(\ell,\ell')} d\tau_4\,\big(\dot x(\tau_4)\cdot\del_{u_4}\big)^{s_4}\,\Pi^{(s_4)}\big(x(\tau_4),u_4;\ell_4,\lambda_4\big) \\
        &\qquad\qquad \times Q^{(s_5)}\int_{\gamma(\ell,\ell')} d\tau_5\,\big(\dot x(\tau_5)\cdot\del_{u_5}\big)^{s_5}\,\Pi^{(s_5)}\big(x(\tau_5),u_5;\ell_5,\lambda_5\big) \ .
    \end{split}
\end{align}
Diagram no. 22: $\vcenter{\includegraphics[scale=0.2]{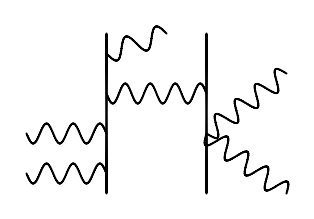}}$
After identifying the worldlines, this contracts into: $\vcenter{\includegraphics[scale=0.2]{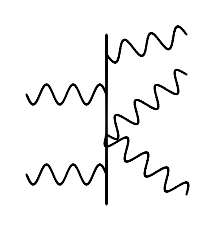}}$
Both diagrams have $5\cdot\binom{4}{2} = 30$ inequivalent orderings of the external legs. For one sample ordering, the contracted diagram reads:
\begin{align}
    \begin{split}
        &\frac{1}{2}\times\frac{N}{4} \int d^3\ell\,d^3\ell'\,G(\ell,\ell')\,Q^{(s_1)}\int_{\gamma(\ell,\ell')} d\tau_1\,\big(\dot x(\tau_1)\cdot\del_{u_1}\big)^{s_1}\,\Pi^{(s_1)}\big(x(\tau_1),u_1;\ell_1,\lambda_1\big) \\
        &\qquad\qquad \times Q^{(s_2)}\int_{\gamma(\ell,\ell')} d\tau_2\,\big(\dot x(\tau_2)\cdot\del_{u_2}\big)^{s_2}\,\Pi^{(s_2)}\big(x(\tau_2),u_2;\ell_2,\lambda_2\big) \\
        &\quad \times \Box_\ell\Box_{\ell'}\,G(\ell,\ell')\,Q^{(s_3)}\int_{\gamma(\ell,\ell')} d\tau_3\,\big(\dot x(\tau_3)\cdot\del_{u_3}\big)^{s_3}\,\Pi^{(s_3)}\big(x(\tau_3),u_3;\ell_3,\lambda_3\big) \\
        &\quad \times \int_{\gamma(\ell,\ell')} d\tau\,V^{(s_4,s_5)}_{\text{new}}\big(\del_{x_4},\del_{u_4};\del_{x_5},\del_{u_5};\dot x(\tau)\big) \\
        &\qquad\qquad\qquad\qquad \times \Pi^{(s_4)}(x_4,u_4;\ell_4,\lambda_4)\,\Pi^{(s_5)}(x_5,u_5;\ell_5,\lambda_5) \Big|_{x_4=x_5=x(\tau)} \ .
    \end{split}
\end{align}
Diagram no. 23: $\vcenter{\includegraphics[scale=0.2]{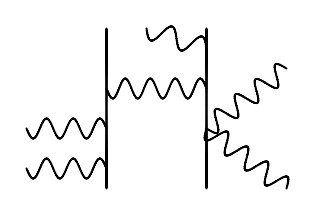}}$
After identifying the worldlines, this contracts into: $\vcenter{\includegraphics[scale=0.2]{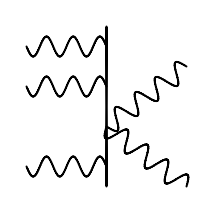}}$
The original diagram has $5\cdot\binom{4}{2} = 30$ inequivalent orderings of the external legs, but the contracted diagram has only $\binom{5}{2} = 10$. Thus, when we sum over inequivalent orderings in the contracted diagram, we should multiply by a factor of $3$.  For one sample ordering, the contracted diagram reads:
\begin{align}
    \begin{split}
        &\frac{3}{2}\times\frac{N}{4} \int d^3\ell\,d^3\ell'\,G(\ell,\ell')\,Q^{(s_1)}\int_{\gamma(\ell,\ell')} d\tau_1\,\big(\dot x(\tau_1)\cdot\del_{u_1}\big)^{s_1}\,\Pi^{(s_1)}\big(x(\tau_1),u_1;\ell_1,\lambda_1\big) \\
        &\qquad\qquad \times Q^{(s_2)}\int_{\gamma(\ell,\ell')} d\tau_2\,\big(\dot x(\tau_2)\cdot\del_{u_2}\big)^{s_2}\,\Pi^{(s_2)}\big(x(\tau_2),u_2;\ell_2,\lambda_2\big) \\
        &\qquad\qquad \times Q^{(s_3)}\int_{\gamma(\ell,\ell')} d\tau\,\big(\dot x(\tau)\cdot\del_{u}\big)^{s_3}\,\Pi^{(s_3)}\big(x(\tau),u;\ell_3,\lambda_3\big) \\
        &\quad \times \Box_\ell\Box_{\ell'}\,G(\ell,\ell') \int_{\gamma(\ell,\ell')} d\tau\,V^{(s_4,s_5)}_{\text{new}}\big(\del_{x_4},\del_{u_4};\del_{x_5},\del_{u_5};\dot x(\tau)\big) \\
        &\qquad\qquad\qquad\qquad \times \Pi^{(s_4)}(x_4,u_4;\ell_4,\lambda_4)\,\Pi^{(s_5)}(x_5,u_5;\ell_5,\lambda_5) \Big|_{x_4=x_5=x(\tau)} \ .
    \end{split}
\end{align}
Diagram no. 24: $\vcenter{\includegraphics[scale=0.2]{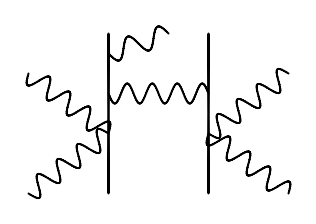}}$
After identifying the worldlines, this contracts into: $\vcenter{\includegraphics[scale=0.2]{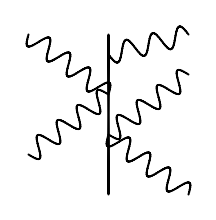}}$
Both diagrams have $5\cdot\binom{4}{2} = 30$ inequivalent orderings of the external legs. For one sample ordering, the contracted diagram reads:
\begin{align}
    \begin{split}
        &\frac{1}{2}\times\frac{N}{4} \int d^3\ell\,d^3\ell'\,G(\ell,\ell') \int_{\gamma(\ell,\ell')} d\tau_{\text{left}}\,
        V^{(s_1,s_2)}_{\text{new}}\big(\del_{x_1},\del_{u_1};\del_{x_2},\del_{u_2};\dot x(\tau_{\text{left}})\big) \\
        &\qquad\qquad\qquad\qquad \times \Pi^{(s_1)}(x_1,u_1;\ell_1,\lambda_1)\,\Pi^{(s_2)}(x_2,u_2;\ell_2,\lambda_2) \Big|_{x_1=x_2=x(\tau_{\text{left}})} \\
        &\quad \times \Box_\ell\Box_{\ell'}\,G(\ell,\ell')\,Q^{(s_3)}\int_{\gamma(\ell,\ell')} d\tau\,\big(\dot x(\tau)\cdot\del_{u}\big)^{s_3}\,\Pi^{(s_3)}\big(x(\tau),u;\ell_3,\lambda_3\big) \\ 
        &\quad \times \int_{\gamma(\ell,\ell')} d\tau_{\text{right}}\,V^{(s_4,s_5)}_{\text{new}}\big(\del_{x_4},\del_{u_4};\del_{x_5},\del_{u_5};\dot x(\tau_{\text{right}})\big) \\
        &\qquad\qquad\qquad\qquad \times \Pi^{(s_4)}(x_4,u_4;\ell_4,\lambda_4)\,\Pi^{(s_5)}(x_5,u_5;\ell_5,\lambda_5) \Big|_{x_4=x_5=x(\tau_{\text{right}})} \ .
    \end{split}
\end{align}

\end{widetext}

\end{document}